%% file: main.tex
\pdfoutput=1
\documentclass[onecolumn,12pt,a4paper]{article}

\usepackage{ifthen} % for conditional statements
\newboolean{pdflatex}
\setboolean{pdflatex}{true} % False for eps figures 

\newboolean{articletitles}
\setboolean{articletitles}{true} % False removes titles in references

\newboolean{uprightparticles}
\setboolean{uprightparticles}{false} %True for upright particle symbols

\def\paperauthors{LHCb collaboration}
\def\paperasciititle{First determination of the spin-parities of the Xic3055 baryons} % Set ASCII title here !! MAKE sure it's only ASCII characters !! 
\def\papertitle{First determination of the spin-parity of $\Xic(3055)^{+,0}$ baryons}
\def\paperkeywords{{High Energy Physics}, {LHCb}, {Heavy Baryon Spectroscopy}} % Comma separated list
\def\papercopyright{\the\year\ CERN for the benefit of the LHCb collaboration} % new since 9/Apr/2018
\def\paperlicence{CC BY 4.0 licence}
\def\paperlicenceurl{https://creativecommons.org/licenses/by/4.0/}

\newif\ifEnableSectionTOCLinks
\EnableSectionTOCLinksfalse % deactivated

\input{preamble}
\usepackage{feynmp-auto}
\usepackage{cleveref}
\begin{document}

%%%%%%%%%%%%%%%%%%%%%%%%%
%%%%% Title     %%%%%%%%%
%%%%%%%%%%%%%%%%%%%%%%%%%
\renewcommand{\thefootnote}{\fnsymbol{footnote}}
\setcounter{footnote}{1}

% %%%%%%% CHOOSE TITLE PAGE--------
\onecolumn
\input{title-LHCb-PAPER}

% %%%%%%%%%%%%% ---------

\renewcommand{\thefootnote}{\arabic{footnote}}
\setcounter{footnote}{0}

%%%%%%%%%%%%%%%%%%%%%%%%%%%%%%%%
%%%%%  Table of Content   %%%%%%
%%%%%%%%%%%%%%%%%%%%%%%%%%%%%%%%
%%%% Uncomment if desired
% \tableofcontents
\cleardoublepage

%%%%%%%%%%%%%%%%%%%%%%%%%
%%%%% Main text %%%%%%%%%
%%%%%%%%%%%%%%%%%%%%%%%%%

\pagestyle{plain} % restore page numbers for the main text
\setcounter{page}{1}
\pagenumbering{arabic}

%% Uncomment during review phase. 
%% Comment before a final submission.
% \linenumbers

%% This is the main body
%% It is useful to have a single file so comments are not missed in overleaf.
\input{body}
\input{acknowledgements}

\input{supplemental-journal}

\addcontentsline{toc}{section}{References}
%\setboolean{inbibliography}{true}
\bibliographystyle{LHCb}
\bibliography{main,standard,LHCb-PAPER,LHCb-CONF,LHCb-DP,LHCb-TDR}

\newpage
\input{Authorship_LHCb-PAPER-2024-018}

\end{document}

%% file: preamble.tex
\usepackage[top=1in, bottom=1.25in, left=1in, right=1in]{geometry}

\usepackage{microtype}
\usepackage{lineno}  % for line numbering during review
\usepackage{xspace} % To avoid problems with missing or double spaces after
\usepackage{caption} %these three command get the figure and table captions automatically small

\usepackage{graphicx}  % to include figures (can also use other packages)
\usepackage{color}
\usepackage{colortbl}
\graphicspath{{./figs/}} % Make Latex search fig subdir for figures
\usepackage{amsmath} % Adds a large collection of math symbols
\usepackage{amssymb}
\usepackage{amsfonts}
\usepackage{upgreek} % Adds in support for greek letters in roman typeset

\newcommand*\patchAmsMathEnvironmentForLineno[1]{%
\expandafter\let\csname old#1\expandafter\endcsname\csname #1\endcsname
\expandafter\let\csname oldend#1\expandafter\endcsname\csname
end#1\endcsname
 \renewenvironment{#1}%
   {\linenomath\csname old#1\endcsname}%
   {\csname oldend#1\endcsname\endlinenomath}%
}
\newcommand*\patchBothAmsMathEnvironmentsForLineno[1]{%
  \patchAmsMathEnvironmentForLineno{#1}%
  \patchAmsMathEnvironmentForLineno{#1*}%
}
\AtBeginDocument{%
\patchBothAmsMathEnvironmentsForLineno{equation}%
\patchBothAmsMathEnvironmentsForLineno{align}%
\patchBothAmsMathEnvironmentsForLineno{flalign}%
\patchBothAmsMathEnvironmentsForLineno{alignat}%
\patchBothAmsMathEnvironmentsForLineno{gather}%
\patchBothAmsMathEnvironmentsForLineno{multline}%
\patchBothAmsMathEnvironmentsForLineno{eqnarray}%
}

\usepackage{hyperxmp}

\usepackage[pdftex,
            pdfauthor={\paperauthors},
            pdftitle={\paperasciititle},
            pdfkeywords={\paperkeywords},
            pdfcopyright={Copyright (C) \papercopyright},
            pdflicenseurl={\paperlicenceurl}]{hyperref}
\usepackage[colorinlistoftodos,textsize=scriptsize]{todonotes}

\usepackage[bottom,flushmargin,hang,multiple]{footmisc}

\usepackage[all]{hypcap} % Internal hyperlinks to floats.

\input{lhcb-symbols-def} % Add in the predefined LHCb symbols

\hypersetup{
  colorlinks   = true, %Colours links instead of ugly boxes
  urlcolor     = blue, %Colour for external hyperlinks
  linkcolor    = blue, %Colour of internal links
  citecolor    = red   %Colour of citations
}

\ifEnableSectionTOCLinks
    \usepackage[explicit]{titlesec} % to change headings
    
    \let\oldcontentsline\contentsline
    \renewcommand

    \titleformat{\section}{\normalfont\Large\bf}{\hyperlink{tocsection.\thesection}{{\thesection} \parbox[t]{\dimexpr\textwidth-1pc}{#1}}}{1pc}{}

    \titleformat{\subsection}{\normalfont\bf}{\hyperlink{tocsubsection.\thesubsection}{{\thesubsection} \parbox[t]{\dimexpr\textwidth-1pc}{#1}}}{1pc}{}

    \titleformat{name=\section,numberless}[display]{}{}{0pt}{\normalfont\Huge\bfseries #1}
\fi

\usepackage{cite} % Allows for ranges in citations
\usepackage{mciteplus}

%% file: lhcb-symbols-def.tex
\usepackage{xspace} 
\usepackage{upgreek}

\def\lhcb   {\mbox{LHCb}\xspace}

\def\babar  {\mbox{BaBar}\xspace}
\def\belle  {\mbox{Belle}\xspace}

\def\besiii {\mbox{BESIII}\xspace}

\def\lhc    {\mbox{LHC}\xspace}

\def\MagUp {\mbox{\em Mag\kern -0.05em Up}\xspace}

\ifthenelse{\boolean{uprightparticles}}%
{

 \def\Ppi         {\ensuremath{\uppi}\xspace}

 \def\PDelta      {\ensuremath{\Delta}\xspace}                 
 \def\PXi         {\ensuremath{\Xi}\xspace}                 
 \def\PLambda     {\ensuremath{\Lambda}\xspace}                 
 \def\PSigma      {\ensuremath{\Sigma}\xspace}                 
 \def\POmega      {\ensuremath{\Omega}\xspace}                 
 \def\PUpsilon    {\ensuremath{\Upsilon}\xspace}
 \let\oldPi\Pi
 \def\PPi         {\ensuremath{\oldPi}\xspace}

 \def\PB      {\ensuremath{\mathrm{B}}\xspace}                 
 \def\PD      {\ensuremath{\mathrm{D}}\xspace}                 

 \def\PK      {\ensuremath{\mathrm{K}}\xspace}                 
 \def\Pb      {\ensuremath{\mathrm{b}}\xspace}                 
 \def\Pc      {\ensuremath{\mathrm{c}}\xspace}

 \def\Pp      {\ensuremath{\mathrm{p}}\xspace}                 

 \def\Ps      {\ensuremath{\mathrm{s}}\xspace}

 \def\thebaroffset{0.0em}
}
{

 \def\Ppi         {\ensuremath{\pi}\xspace}

 \mathchardef\PDelta="7101
 \mathchardef\PXi="7104
 \mathchardef\PLambda="7103
 \mathchardef\PSigma="7106
 \mathchardef\POmega="710A
 \mathchardef\PUpsilon="7107
 \mathchardef\PPi="7105
 \def\PB      {\ensuremath{B}\xspace}                 
 \def\PD      {\ensuremath{D}\xspace}                 

 \def\PK      {\ensuremath{K}\xspace}                 
 \def\Pb      {\ensuremath{b}\xspace}                 
 \def\Pc      {\ensuremath{c}\xspace}

 \def\Pp      {\ensuremath{p}\xspace}                 

 \def\Ps      {\ensuremath{s}\xspace}

 \def\thebaroffset{0.18em}
}
\newcommand{\offsetoverline}[2][\thebaroffset]{\kern #1\overline{\kern -#1 #2}}%

\makeatletter
\ifcase \@ptsize \relax% 10pt
  \newcommand{\miniscule}{\@setfontsize\miniscule{4}{5}}% \tiny: 5/6
\or% 11pt
  \newcommand{\miniscule}{\@setfontsize\miniscule{5}{6}}% \tiny: 6/7
\or% 12pt
  \newcommand{\miniscule}{\@setfontsize\miniscule{5}{6}}% \tiny: 6/7
\fi
\makeatother

\DeclareRobustCommand{\optbar}[1]{\shortstack{{\miniscule (\rule[.5ex]{1.25em}{.18mm})}
  \\ [-.7ex] $#1$}}

\def\squark    {{\ensuremath{\Ps}}\xspace}

\def\cquark    {{\ensuremath{\Pc}}\xspace}

\def\bquark    {{\ensuremath{\Pb}}\xspace}

\def\pion   {{\ensuremath{\Ppi}}\xspace}

\def\pip    {{\ensuremath{\pion^+}}\xspace}
\def\pim    {{\ensuremath{\pion^-}}\xspace}

\def\kaon    {{\ensuremath{\PK}}\xspace}
\def\KorKbar {\kern \thebaroffset\optbar{\kern -\thebaroffset \PK}{}\xspace}

\def\Km      {{\ensuremath{\kaon^-}}\xspace}

\def\D       {{\ensuremath{\PD}}\xspace}

\def\DorDbar {\kern \thebaroffset\optbar{\kern -\thebaroffset \PD}\xspace}
\def\Dz      {{\ensuremath{\D^0}}\xspace}

\def\Dp      {{\ensuremath{\D^+}}\xspace}
\def\Dm      {{\ensuremath{\D^-}}\xspace}

\def\DpDm    {\ensuremath{\Dp {\kern -0.16em \Dm}}\xspace}

\def\B       {{\ensuremath{\PB}}\xspace}

\def\BorBbar {\kern \thebaroffset\optbar{\kern -\thebaroffset \PB}\xspace}

\def\Bd      {{\ensuremath{\B^0}}\xspace}

\def\BdorBdbar {\kern \thebaroffset\optbar{\kern -\thebaroffset \Bd}\xspace}

\def\Bs      {{\ensuremath{\B^0_\squark}}\xspace}

\def\BsorBsbar {\kern \thebaroffset\optbar{\kern -\thebaroffset \Bs}\xspace}

\def\Y#1S{\ensuremath{\PUpsilon{(#1S)}}\xspace}

\def\proton      {{\ensuremath{\Pp}}\xspace}

\def\Lz          {{\ensuremath{\PLambda}}\xspace}

\def\LorLbar     {\kern \thebaroffset\optbar{\kern -\thebaroffset \PLambda}\xspace}

\def\Sigmares    {{\ensuremath{\PSigma}}\xspace}
\def\Sigmaz      {{\ensuremath{\Sigmares{}^0}}\xspace}

\def\Xires       {{\ensuremath{\PXi}}\xspace}

\def\Sigmac       {{\ensuremath{\Sigmares_\cquark}}\xspace}

\def\Sigmacpp      {{\ensuremath{\Sigmares_\cquark^{++}}}\xspace}

\def\Xic         {{\ensuremath{\Xires_\cquark}}\xspace}

\def\Xicp        {{\ensuremath{\Xires^+_\cquark}}\xspace}

\def\Lb           {{\ensuremath{\Lz^0_\bquark}}\xspace}

\def\Xib          {{\ensuremath{\Xires_\bquark}}\xspace}
\def\Xibz         {{\ensuremath{\Xires^0_\bquark}}\xspace}
\def\Xibm         {{\ensuremath{\Xires^-_\bquark}}\xspace}

\def\BF         {{\ensuremath{\mathcal{B}}}\xspace}

\newcommand{\decay}[2]{\ensuremath{#1\!\to #2}\xspace} 

\def\to                 {\ensuremath{\rightarrow}\xspace}

\def\AT#1     {\ensuremath{A_{\mathrm{T}}^{#1}}\xspace}           % 2

\def\C#1      {\ensuremath{\mathcal{C}_{#1}}\xspace}                       % 9
\def\Cp#1     {\ensuremath{\mathcal{C}_{#1}^{'}}\xspace}                    % 7
\def\Ceff#1   {\ensuremath{\mathcal{C}_{#1}^{\mathrm{(eff)}}}\xspace}        % 9  
\def\Cpeff#1  {\ensuremath{\mathcal{C}_{#1}^{'\mathrm{(eff)}}}\xspace}       % 7
\def\Ope#1    {\ensuremath{\mathcal{O}_{#1}}\xspace}                       % 2
\def\Opep#1   {\ensuremath{\mathcal{O}_{#1}^{'}}\xspace}                    % 7

\newcommand{\aunit}[1]{\ensuremath{\text{\,#1}}}       
\newcommand{\tev}{\aunit{Te\kern -0.1em V}\xspace}
\newcommand{\gev}{\aunit{Ge\kern -0.1em V}\xspace}
\newcommand{\mev}{\aunit{Me\kern -0.1em V}\xspace}
\newcommand{\kev}{\aunit{ke\kern -0.1em V}\xspace}
\newcommand{\ev}{\aunit{e\kern -0.1em V}\xspace}
 
\newcommand{\mevc}{\ensuremath{\aunit{Me\kern -0.1em V\!/}c}\xspace}
\newcommand{\gevc}{\ensuremath{\aunit{Ge\kern -0.1em V\!/}c}\xspace}
\newcommand{\mevcc}{\ensuremath{\aunit{Me\kern -0.1em V\!/}c^2}\xspace}
\newcommand{\gevcc}{\ensuremath{\aunit{Ge\kern -0.1em V\!/}c^2}\xspace}
\def\fb   {\ensuremath{\aunit{fb}}\xspace}
\def\invfb   {\ensuremath{\fb^{-1}}\xspace}

\newcommand{\stat}{\aunit{(stat)}\xspace}
\newcommand{\syst}{\aunit{(syst)}\xspace}

\def\deriv {\ensuremath{\mathrm{d}}}

\def\gsim{{~\raise.15em\hbox{$>$}\kern-.85em
          \lower.35em\hbox{$\sim$}~}\xspace}
\def\lsim{{~\raise.15em\hbox{$<$}\kern-.85em
          \lower.35em\hbox{$\sim$}~}\xspace}

\def\sPlot{\mbox{\em sPlot}\xspace}

\def\sqs   {\ensuremath{\protect\sqrt{s}}\xspace}

\def\tell1  {TELL1\xspace}
\def\ukl1   {UKL1\xspace}

\newcommand{\lhcborcid}[1]{\href{https://orcid.org/#1}{\hspace*{0.1em}\raisebox{-0.45ex}{\includegraphics[width=1em]{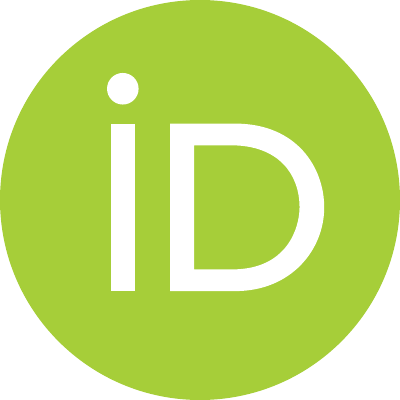}}}}

%% file: title-LHCb-PAPER.tex
% ===============================================================================
% Purpose: LHCb-PAPER journal paper title page template
% Author: 
% Created on: 2010-09-25
% ===============================================================================

%%%%%%%%%%%%%%%%%%%%%%%%%
%%%%%  TITLE PAGE  %%%%%%
%%%%%%%%%%%%%%%%%%%%%%%%%
\begin{titlepage}
\pagenumbering{roman}

% Header ---------------------------------------------------
\vspace*{-1.5cm}
\centerline{\large EUROPEAN ORGANIZATION FOR NUCLEAR RESEARCH (CERN)}
\vspace*{1.5cm}
\noindent
\begin{tabular*}{\linewidth}{lc@{\extracolsep{\fill}}r@{\extracolsep{0pt}}}
\ifthenelse{\boolean{pdflatex}}% Logo format choice
{\vspace*{-1.5cm}\mbox{\!\!\!\includegraphics[width=.14\textwidth]{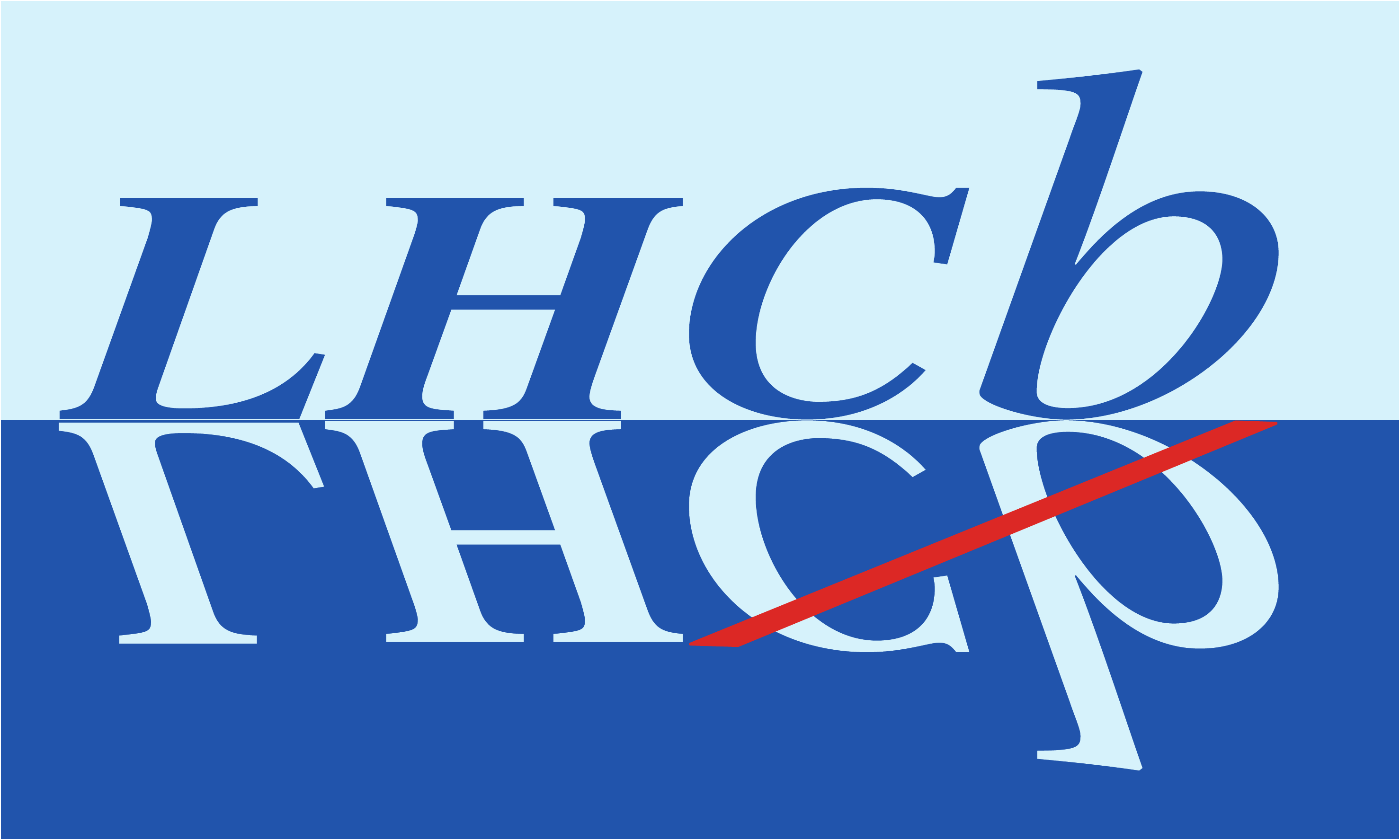}} & &}%
{\vspace*{-1.2cm}\mbox{\!\!\!\includegraphics[width=.12\textwidth]{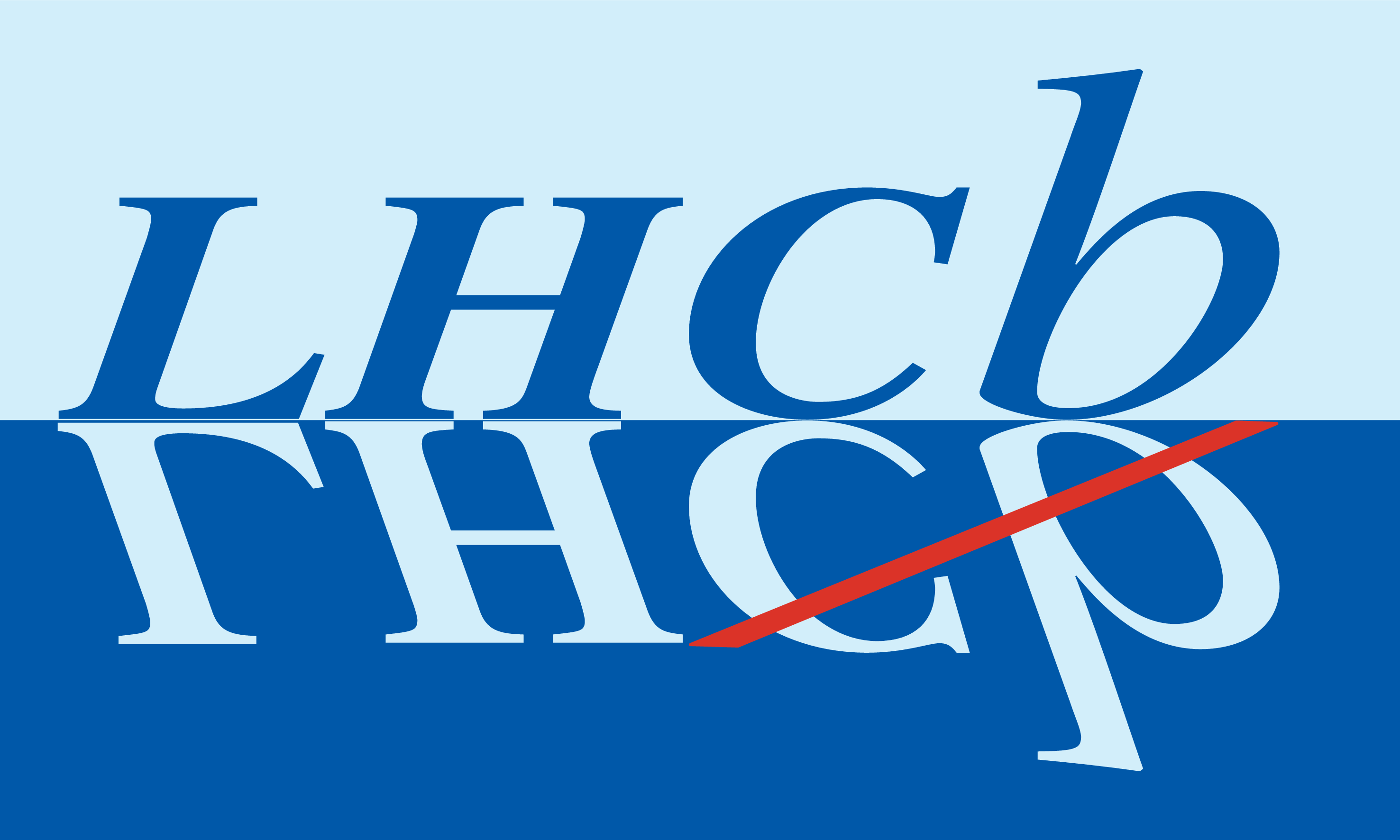}} & &}%
\\
 & & CERN-EP-2024-215 \\  % ID 
 & & LHCb-PAPER-2024-018 \\  % ID 
 & & \today \\ % or \today
 & & \\
% not in paper \hline
\end{tabular*}

\vspace*{4.0cm}

% Title --------------------------------------------------
{\normalfont\bfseries\boldmath\huge
\begin{center}
% DO NOT EDIT HERE. Instead edit macro in main.tex to keep metadata correct
  \papertitle 
\end{center}
}

\vspace*{2.0cm}

% Authors -------------------------------------------------
\begin{center}
%In the footnote, replace 'paper' by 'Letter' in case of submission to PRL or PLB 
% Edit macro in main.tex to keep metadata correct
\paperauthors\footnote{Authors are listed at the end of this paper.}
\end{center}

\vspace{\fill}

% Abstract -----------------------------------------------
\begin{abstract}
%% Plain latex version
The ${{\Xi}_{b}^{0(-)}\to{\Xi}_{c}(3055)^{+(0)}(\to D^{+(0)}{\Lambda}){\pi}^{-}}$ decay chains are observed, and the spin-parity of ${\Xi}_{c}(3055)^{+(0)}$ baryons is determined for the first time. 
The measurement is performed using proton-proton collision data at a center-of-mass energy of $\sqrt{s}=13\,\text{TeV}$, corresponding to an integrated luminosity of $5.4\,\text{fb}^{-1}$, recorded by the~$\text{LHCb}$ experiment between 2016 and 2018.
The spin-parity of the ${\Xi}_{c}(3055)^{+(0)}$ baryons is determined to be $3/2^{+}$ with a significance of more than $6.5\sigma$ ($3.5\sigma$) compared to all other tested hypotheses.
The up-down asymmetries of the ${{\Xi}_{b}^{0(-)}\to{\Xi}_{c}(3055)^{+(0)}{\pi}^{-}}$ transitions are measured to be 
$-0.92\pm0.10\pm0.05$ ($-0.92\pm0.16\pm0.22$), consistent with maximal parity violation, where the first uncertainty is statistical and the second is systematic.
These results support the hypothesis that the ${\Xi}_{c}(3055)^{+(0)}$ baryons correspond to the first $D$-wave $\lambda$-mode excitation of the ${\Xi}_{c}$ flavor triplet.
%% With LHCb symbols
% The ${{\Xires}_{b}^{0(-)}\to{\Xires}_{c}(3055)^{+(0)}(\to D^{+(0)}{\Lz}){\pi}^{-}}$ decay chains are observed, and the spin-parity of ${\Xires}_{c}(3055)^{+(0)}$ baryons is determined for the first time. 
% The measurement is performed using proton-proton collision data at a center-of-mass energy of $\sqrt{s}=13\,\text{TeV}$, corresponding to an integrated luminosity of $5.4\,\text{fb}^{-1}$, recorded by the~$\text{LHCb}$ experiment between 2016 and 2018.
% The spin-parity of the ${\Xires}_{c}(3055)^{+(0)}$ baryons is determined to be $3/2^{+}$ with a significance of more than $6.5\sigma$ ($3.5\sigma$) compared to all other tested hypotheses.
% The up-down asymmetries of the ${{\Xires}_{b}^{0(-)}\to{\Xires}_{c}(3055)^{+(0)}{\pi}^{-}}$ transitions are measured to be 
% $-0.92\pm0.10\pm0.05$ ($-0.92\pm0.16\pm0.22$), consistent with maximal parity violation, where the first uncertainty is statistical and the second is systematic.
% These results support the hypothesis that the ${\Xires}_{c}(3055)^{+(0)}$ baryons correspond to the first $D$-wave $\lambda$-mode excitation of the ${\Xires}_{c}$ flavor triplet.
\end{abstract}

\vspace*{2.0cm}

\begin{center}
  Published in Physical Review Letters 134 (2025) 081901
  % JHEP /
  % Phys.~Rev.~D /
  % 
%  Phys.~Lett.~B /
  % Eur.~Phys.~J.~C /
  %  Nucl.~Phys.~B /
  % Chin.~Phys.~C /
  % Nature~Physics /
  % sciPost~Physics /
  % J. Instr. /
  % Instruments 
\end{center}

\vspace{\fill}

{\footnotesize 
% Edit macro in main.tex to keep metadata correct
\centerline{\copyright~\papercopyright. \href{\paperlicenceurl}{\paperlicence}.}}
\vspace*{2mm}

\end{titlepage}

%%%%%%%%%%%%%%%%%%%%%%%%%%%%%%%%
%%%%%  EOD OF TITLE PAGE  %%%%%%
%%%%%%%%%%%%%%%%%%%%%%%%%%%%%%%%

%  empty page follows the title page ----
\newpage
\setcounter{page}{2}
\mbox{~}
%\newpage
%
%% Author List ----------------------------
%%  You need to get a new author list!
% \input{LHCb_authorlist.tex}
%
%The author list for journal publications is provided by the Membership Committee shortly after 'approval to go to paper' has been given.
%%It will be made available on the page
%%\verb!http://www.physik.uzh.ch/~strauman/forMemCo/LHCb-PAPER-XXXX-XXX/! .
%It will be sent to you by email shortly after a paper number has beens assigned.
%The author list should be included already at first circulation, 
%to allow new members of the collaboration to verify whether they have been included correctly.
%Occasionally a misspelled name is corrected or associated institutions become full members.
%In that case, a new author list will be sent to you.
%In case line numbering doesn't work well after including the authorlist, try moving the \verb!\bigskip! after the last author to a separate line.
%
%
%The authorship for Conference Reports should be ``The LHCb
%  collaboration'', with a footnote giving the name(s) of the contact
%  author(s), but without the full list of collaboration names.

%% file: body.tex
Baryons containing a single heavy quark, hereafter referred to as singly heavy baryons, provide an ideal laboratory for studying the complex quark dynamics.
Their structures can be effectively described by the approximation of a heavy quark and a diquark system of light quarks, with the dynamics primarily governed by the diquark degrees of freedom~\cite{Isgur:1991wq}.
Based on the spin-flavor wave function of the diquark, the ground states can be categorized into flavor antisymmetrical triplets, 
denoted as $\bar{3}_F$, 
and flavor symmetrical sextuplets 
denoted as $6_F$, respectively~\cite{Chen:2022asf}.
Excitation can occur either between the two light quarks, known as the $\rho$ mode, or between the heavy quark and the diquark, referred to as the $\lambda$ mode.
Considering the various excitation modes and spin-angular momentum configurations, a rich spectrum of singly-heavy baryons is expected, providing insights into the confinement mechanism of the strong interaction~\cite{Chen:2016spr}.

Numerous excited singly charmed baryons have been observed by \belle~\cite{Belle:2004zjl,Belle:2006edu,Belle:2006xni,Belle:2008yxs,Solovieva:2008fw,Belle:2013htj,Belle:2016tai,Belle:2017ext}, \babar~\cite{BaBar:2006pve,BaBar:2006itc,BaBar:2007zjt}, and \lhcb~\cite{LHCb-PAPER-2017-002,LHCb-PAPER-2021-012, LHCb-PAPER-2022-043} experiments in the last two decades.
While many theoretically allowed states remain undiscovered, most of the observed resonances, including the $\Xic(3055)^{+(0)}$ baryons, have not yet been well established in terms of their excitation nature.
The $\Xic(3055)^+$ baryon, with quark content $csu$, was observed in the $\Sigmacpp \Km$ and $\Dp \Lz$ final states~\cite{BaBar:2007zjt,Belle:2013htj,Belle:2016tai}.\footnote{The inclusion of charge-conjugated processes is implied throughout.}
Its isospin partner, the $\Xic(3055)^0$ baryon with quark content $csd$, was observed in the $\Dz\!\Lz$ final state~\cite{Belle:2016tai}.
A number of possible explanations for the excitation modes have been proposed based on their masses, widths, and strong decay properties.
In Refs.~\cite{Liu:2007ge,Guo:2008he,Liu:2012sj,Chen:2014nyo,PhysRevD.94.114016,Zhao:2016qmh,Chen:2017aqm,Wang:2017vtv,Ye:2017dra,Yao:2018jmc}, the $\Xic(3055)^{+(0)}$ states are interpreted as the $D$-wave orbital angular momentum excitation with possible spin-parity ($J^P$) assignments of $3/2^+$, $5/2^+$ or $7/2^+$.
The second orbital excitation of the $\lambda$ mode is favored over the $\rho$ mode or a combination of both~\cite{Liu:2012sj,Wang:2017vtv}.
The strong decays of the $\Xic(3055)^{+}$ state to the ${\Sigmac^{++} \Km}$ and ${\Dp\Lz}$ final states, studied in the $^{3}P_0$ model, suggest that it may be a $2S$ excitation of the $\Xic$ ($\bar{3}_F$) or $\Xires_c'$ ($6_F$) state, with $J^P=1/2^+$ or $3/2^+$~\cite{Ye:2017dra}.
Hadron molecular states have also been proposed~\cite{Yu:2018yxl} to interpret the $\Xic(3055)^{+(0)}$ baryons, with $J^P=1/2^-$ or $3/2^-$.
Thus, measurements of the spin-parity of the $\Xic(3055)^{+(0)}$ baryons are crucial to pin down their nature and clarify the complicated charm-baryon spectrum.

The spin-parity of the $\Xic(3055)^{+(0)}$ baryons can be studied by exploiting the weak decays $\Xires_b^{0(-)}\to{\Xic(3055)^{+(0)} \pim}$.
In this work, amplitude analyses of ${\Xires_b^{0(-)}\to\Xires_c^{**+(0)}\pim}$ decays are performed, where the $\Xires_c^{**+(0)}$ states refer to excited $\Xic(3055)^{+(0)}$ or $\Xic(3080)^{+(0)}$ baryons and are reconstructed in the ${\D^{+(0)}\Lz}$ final states. 
The spin-parity, masses and widths of the $\Xic(3055)^{+(0)}$ baryons are determined, as well as the up-down asymmetries of the ${\Xires_b^{0(-)}\to \Xic(3055)^{+(0)} \pim}$ transitions, 
which are defined as the relative difference between the decay rates for the up and down helicity states~\cite{Jacob:1959at} of the ${\Xires_b^{0(-)}}$ baryons.
The analysis is performed using proton-proton ($pp$) collision data at a center-of-mass energy of $\sqs=13\tev$, corresponding to an integrated luminosity of about $5.4\invfb$, collected with the~\lhcb detector between 2016 and 2018.

The LHCb detector, designed for the study of particles containing \bquark\ or \cquark\ quarks, is a single-arm forward spectrometer covering the pseudorapidity range $2 < \eta < 5$, described in detail in Refs.~\cite{LHCb-DP-2008-001,LHCb-DP-2014-002}. 
The online event selection for $\Xib$ decays is performed by a trigger~\cite{LHCb-DP-2012-004} which consists of a hardware stage followed by a two-step software stage~\cite{LHCb-TDR-016, LHCb-PROC-2015-018, BBDT, LHCb-DP-2019-001}.
The hardware trigger decision is based on the transverse energy (the fraction of a particle's total energy that lies perpendicular to the beam axis) deposited in the hadronic calorimeter.
The first step of the software trigger requires a single track or a pair of tracks with sufficient transverse momentum and impact parameter, which is defined as the minimum distance of the track relative to the primary $pp$ interaction vertex (PV). In the second step, the presence of a secondary vertex that is well separated from the PV is required.

Simulated samples are used to optimize the selection criteria, parameterize the invariant-mass distributions and characterize the detector resolution and efficiencies. 
These samples are generated using the software described in Refs.~\cite{Sjostrand:2007gs,*Sjostrand:2006za,LHCb-PROC-2010-056,Lange:2001uf,davidson2015photos,Allison:2006ve, *Agostinelli:2002hh,LHCb-PROC-2011-006}. 
In the simulation, the products of the $\Xires_b^{0(-)}$, $\Xires_c^{**+(0)}$, $\D^{+(0)}$ and $\Lz$ decays are generated uniformly over the allowed phase space.

In the offline reconstruction, charged tracks identified as protons, kaons or pions are combined to form ${\Lz\to\proton\pim}$, ${\Dp\to\Km\pip\pip}$ and ${\Dz\to\Km\pip}$ candidate decays. 
The reconstructed $\Lz$, $\Dp$ and $\Dz$ vertices are required to have good quality and be significantly displaced from any PV. 
The invariant masses of $\D^{+(0)}$ and $\Lz$ candidates are required to be within $\pm20\mevcc$ and $\pm6\mevcc$ of the known values~\cite{PDG2024}, respectively.
The $\D^{+(0)}$ and $\Lz$ candidates are combined with an additional $\pim$ track to form $\Xires_b^{0(-)}$ candidates. 
To improve the resolution of the reconstructed $\Xires_b^{0(-)}$ invariant mass, denoted as $m_{\D^{+(0)}\Lz\pim}$, a kinematic fit is applied~\cite{Hulsbergen:2005pu}, constraining the $\D^{+(0)}$ and $\Lz$ invariant masses to their known values and imposing the $\Xires_b^{0(-)}$ momentum to point back to the associated PV.
The multi-layer perceptron (MLP) neural network implemented in the TMVA toolkit~\cite{TMVA4} is utilized to further distinguish the $\Xires_b^{0(-)}$ signal from the combinatorial background.
An MLP classifier is trained for $\Xibz$ and $\Xibm$ decays independently, which combines the kinematic and vertexing information of the $\Xires_b^{0(-)}$ baryon and its decay products.

\begin{figure}[!tb]
    \centering
    \includegraphics[width=0.48\columnwidth]{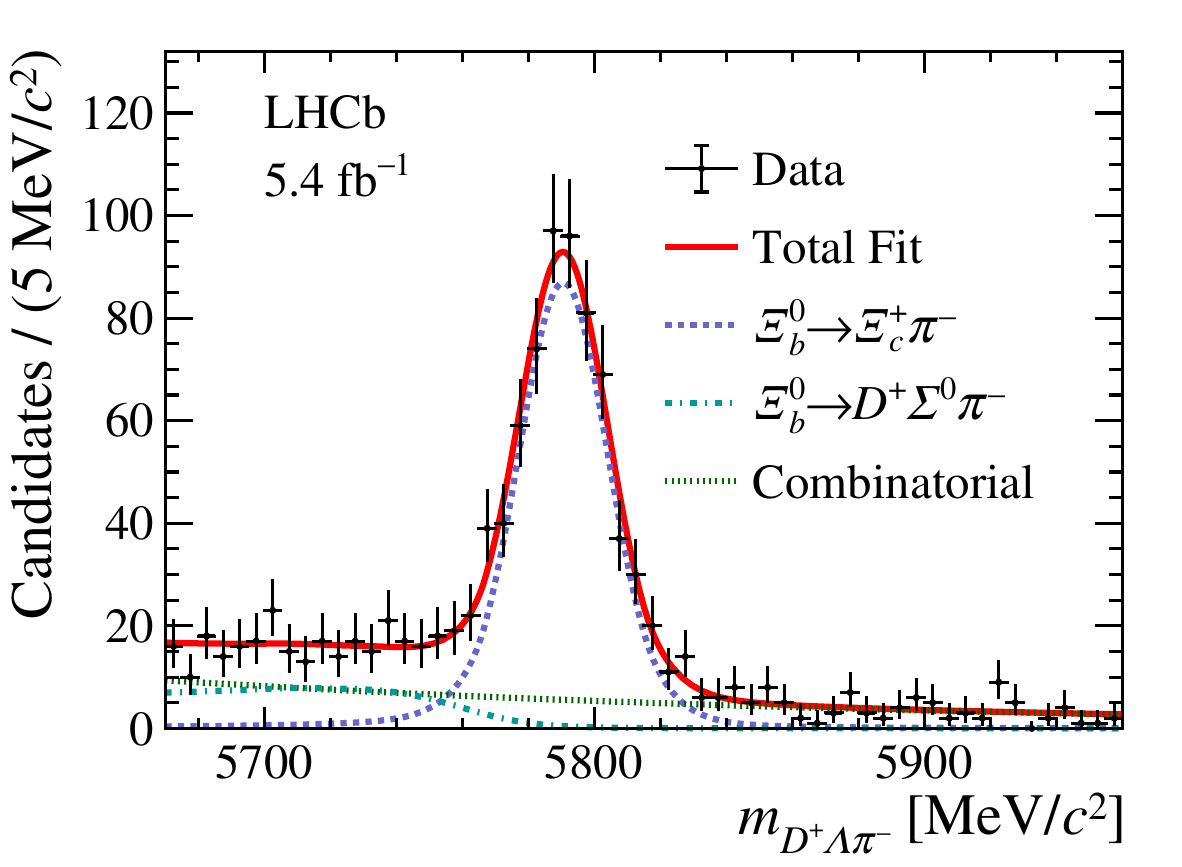}
    \includegraphics[width=0.48\columnwidth]{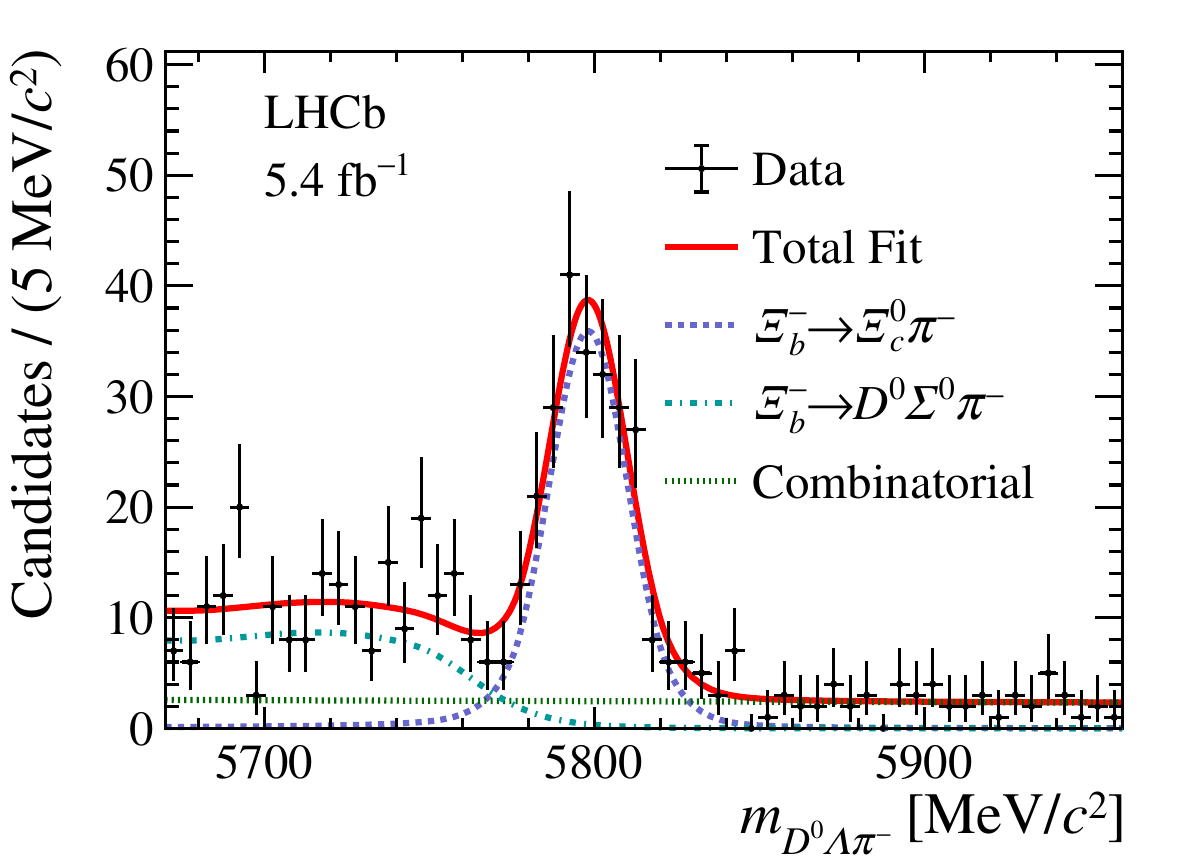}
    \caption{Distributions of the (left) $\Dp\Lz\pim$ and (right) $\Dz\Lz\pim$ invariant mass with the fit results overlaid.}
    \label{fig:massfit_Xib}
\end{figure}

Extended maximum-likelihood fits are performed to the $m_{\D^{+(0)}\Lz\pim}$ distributions to determine the signal yields.
The $\Xires_b^{0(-)}$ signal components are described by the combination of a Gaussian function and a double-sided Crystal Ball function~\cite{Skwarnicki:1986xj}. 
Background due to partially reconstructed ${\decay{\Xires_b^{0(-)}}{\D^{+(0)}\Sigmaz(\to\Lz\gamma)\pim}}$ decays, with the photon missing, is modelled from simulation using a sample generated with a fast parametric method~\cite{Cowan:2016tnm}.
The combinatorial background is described by an exponential function. 
The yields of the signal and these two background components are allowed to vary in the fits. 
The $m_{\D^{+(0)}\Lz\pim}$ distributions and fit results are shown in Fig.~\ref{fig:massfit_Xib}.
The total $\Xires_b^{0(-)}$ yields are measured to be $637\pm31$ ($232\pm19$).
The \sPlot technique~\cite{Pivk:2004ty} is used to assign a weight to each $\Xires_b^{0(-)}$ candidate based on the fit results to subtract the background.

\begin{figure}[!tb]
    \centering
    \includegraphics[width=0.48\columnwidth]{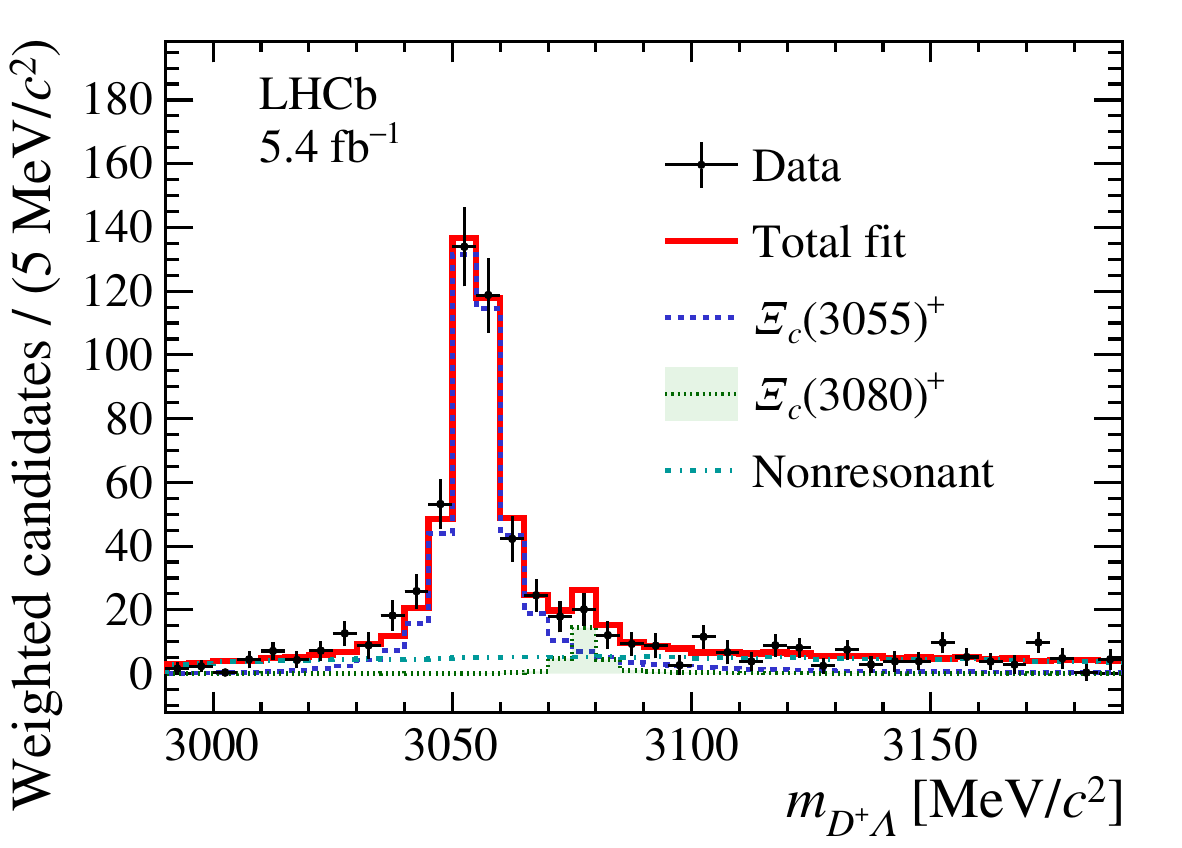}
    \includegraphics[width=0.48\columnwidth]{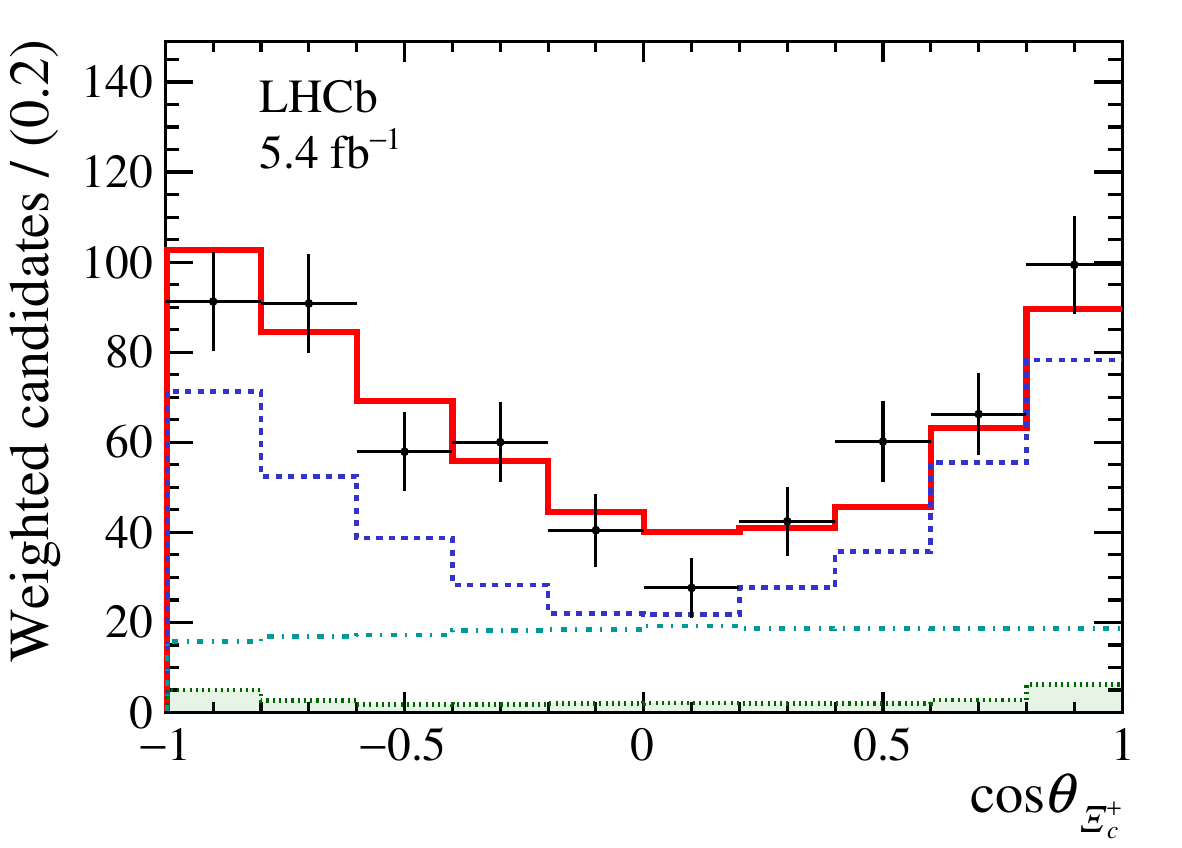}
    \includegraphics[width=0.48\columnwidth]{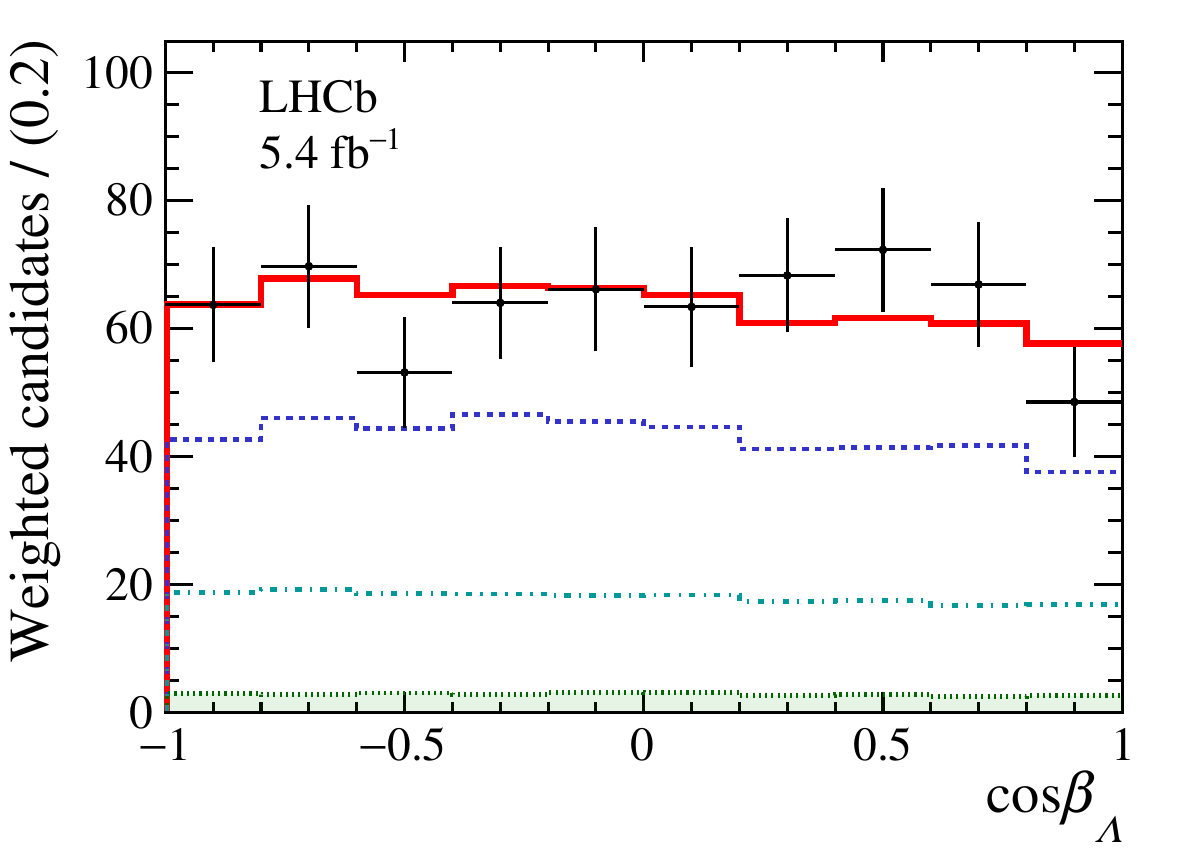}
    \includegraphics[width=0.48\columnwidth]{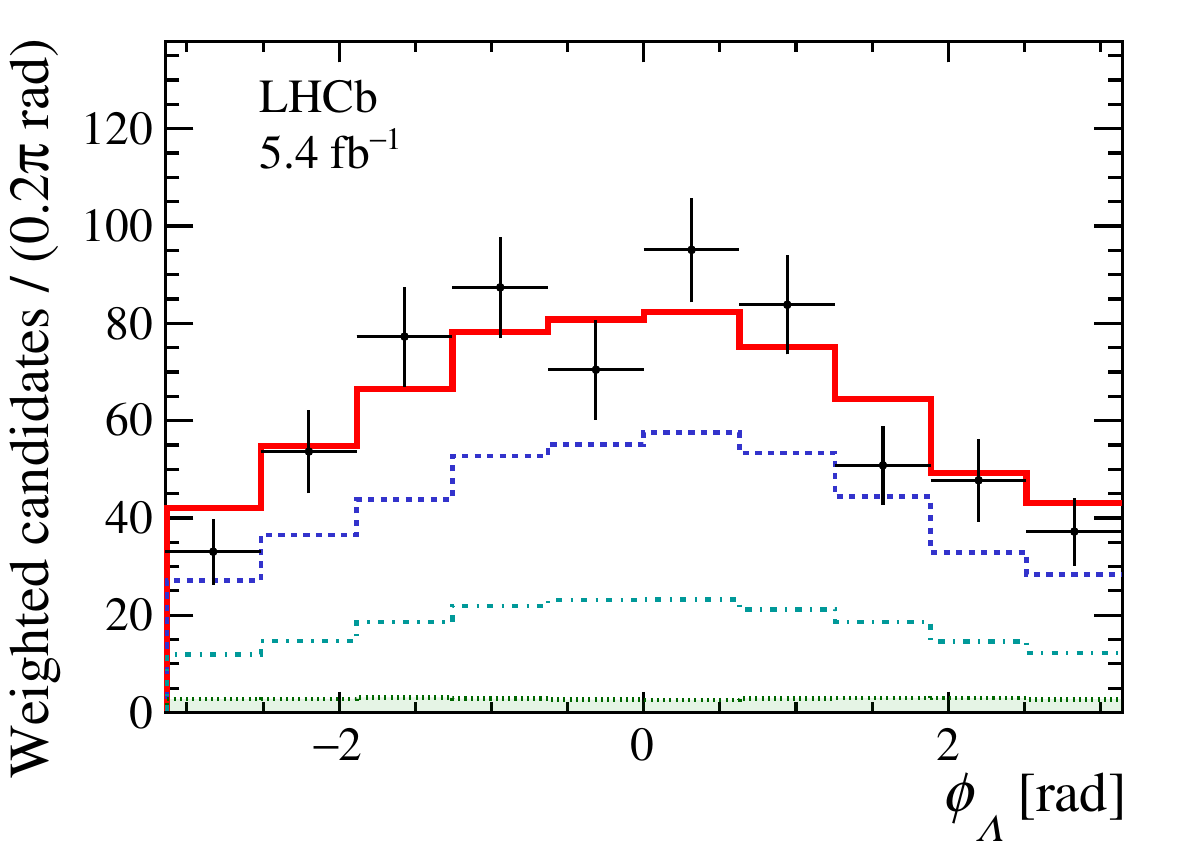}
    \caption{Distributions of the (top left) $\Dp\!\Lz$ invariant mass, (top right) $\Xires_c^{+}$ helicity angle, (bottom left) $\Lz$ helicity angle and (bottom right) azimuthal angle, for the  ${\Xibz\to\Dp\Lz\pim}$ sample.  The projections of the amplitude fit under the spin-parity hypothesis $J^P_{\Xic(3055)^+}=3/2^+$ are overlaid.}
    \label{fig:ampfit_Xibz}
\end{figure}

\begin{figure}[!tb]
    \centering
    \includegraphics[width=0.48\columnwidth]{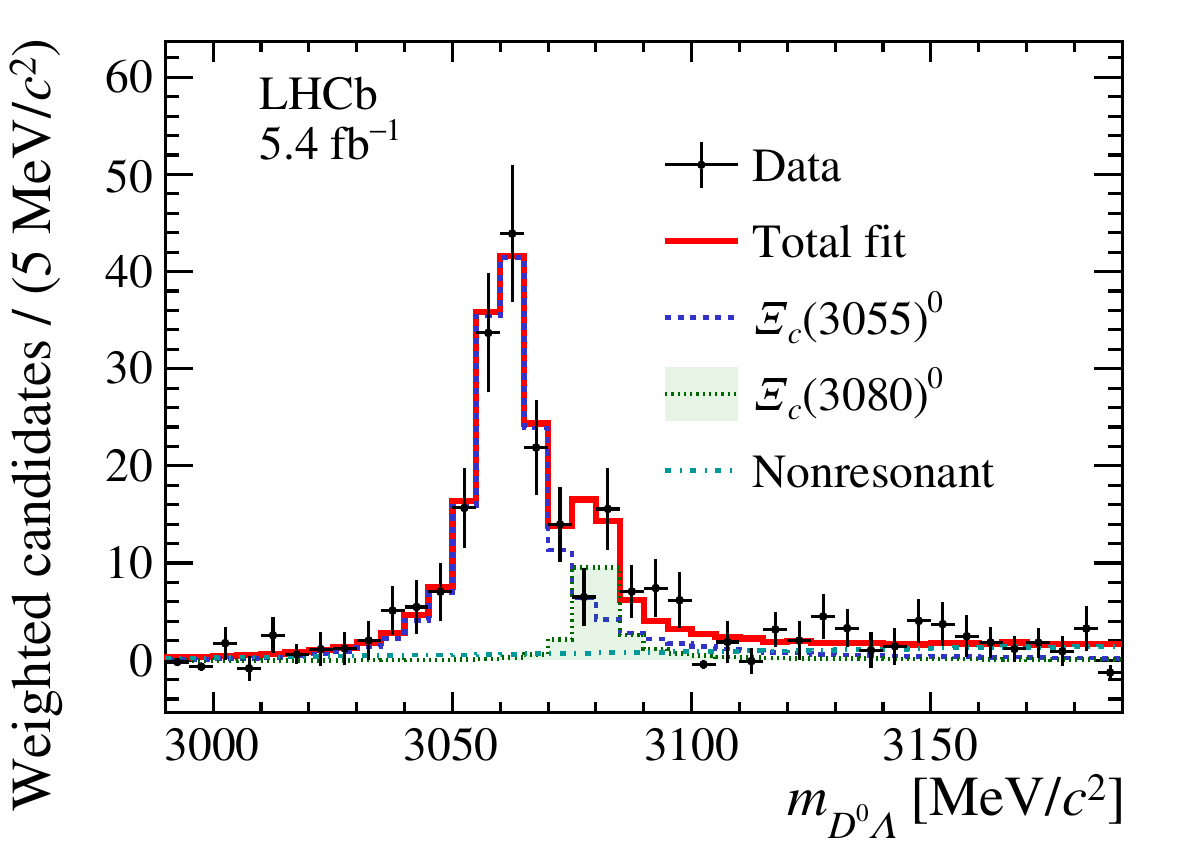}
    \includegraphics[width=0.48\columnwidth]{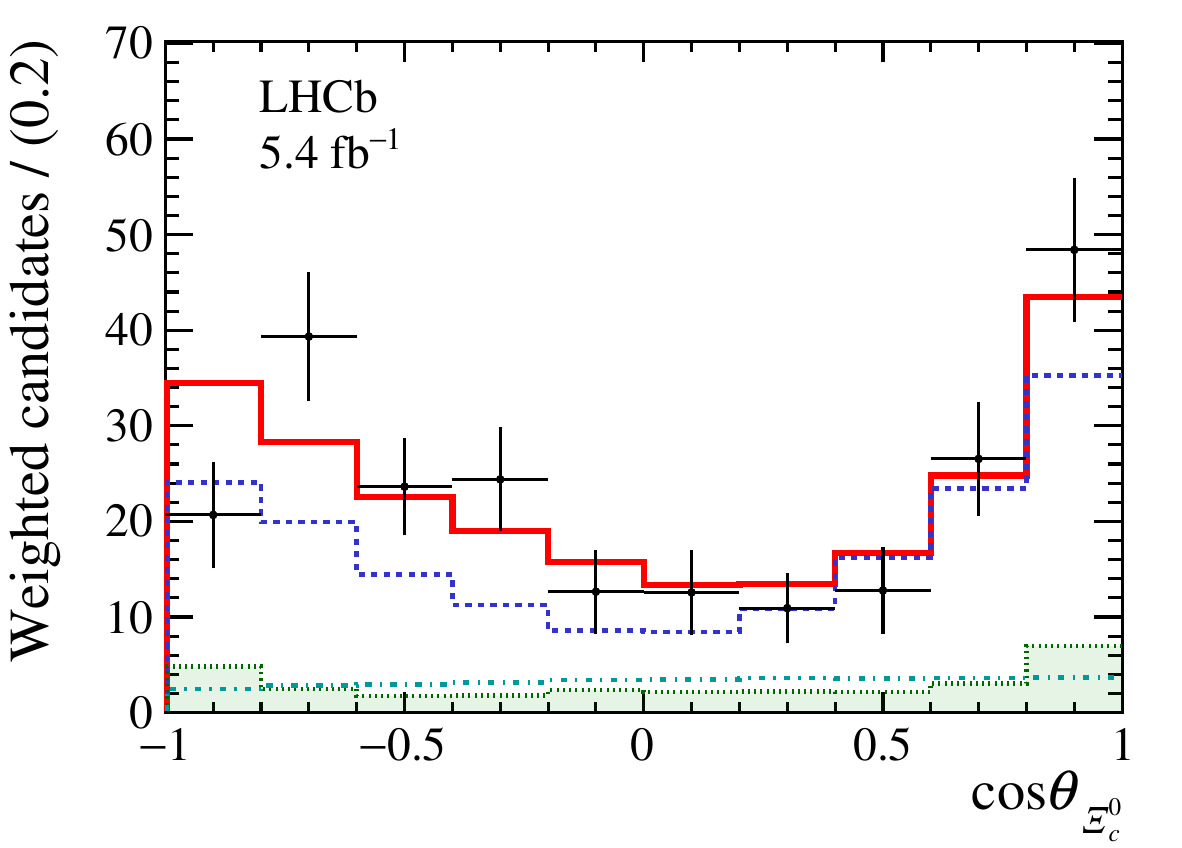}
    \includegraphics[width=0.48\columnwidth]{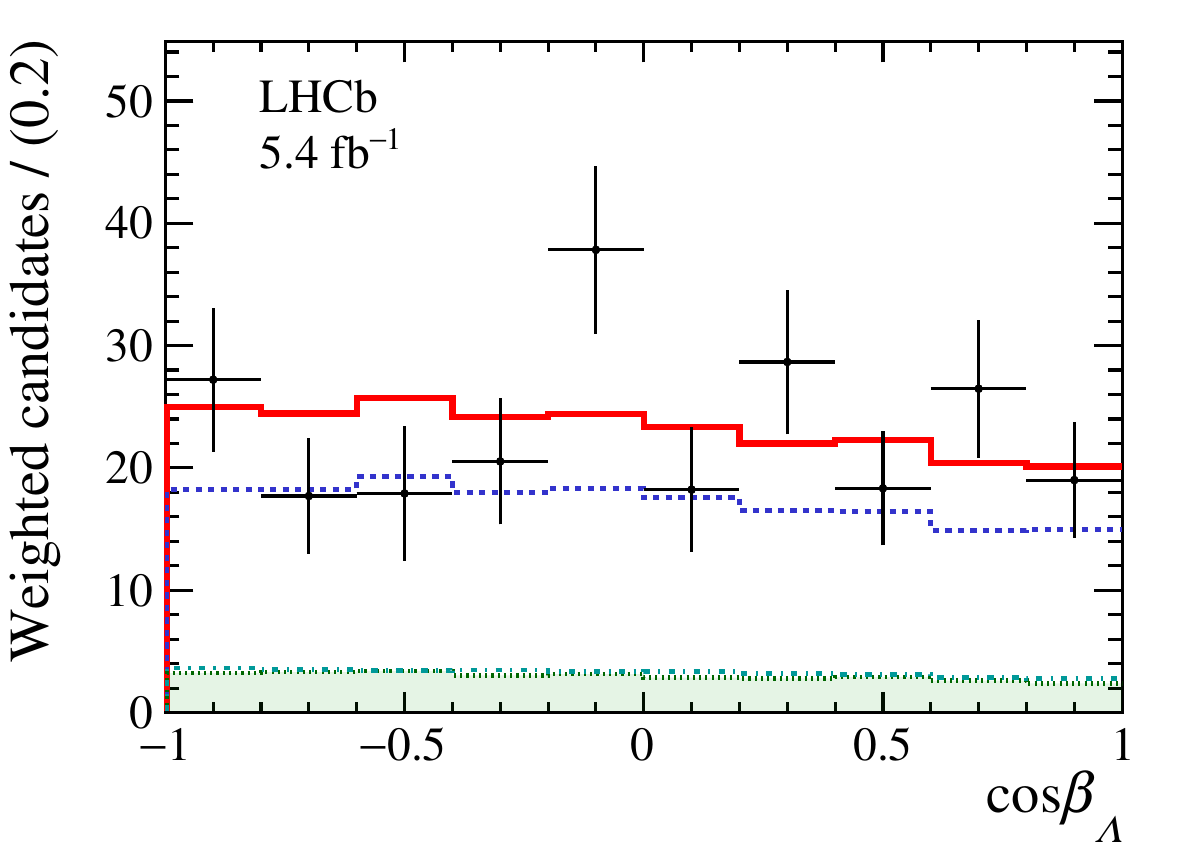}
    \includegraphics[width=0.48\columnwidth]{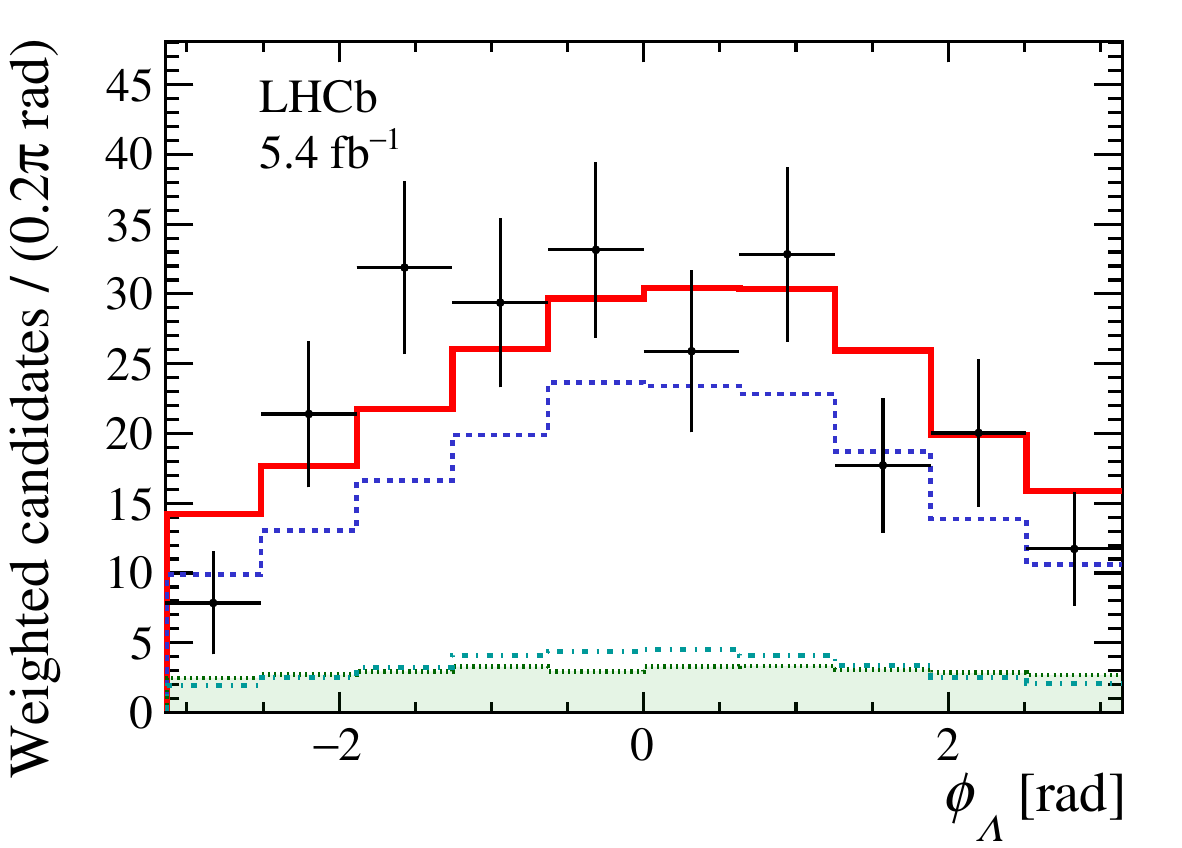}
    \caption{Distributions of the (top left) $\Dz\!\Lz$ invariant mass, (top right) $\Xires_c^{0}$ helicity angle, (bottom left) $\Lz$ helicity angle and (bottom right) azimuthal angle, for the  ${\Xibm\to\Dz\Lz\pim}$ sample.  The projections of the amplitude fit under the spin-parity hypothesis $J^P_{\Xic(3055)^0}=3/2^+$ are overlaid.}
    \label{fig:app_ampfit_Xibm}
\end{figure}

The polarization of $\Lb$ baryons at the~\lhc has been measured to be consistent with zero~\cite{LHCb-PAPER-2020-005}. 
Assuming the $\Xires_b^{0(-)}$ baryons are also produced unpolarized, the \mbox{$\decay{\Xires_b^{0(-)}}{\D^{+(0)}\Lz\pim}$} decay kinematics are fully described by the invariant mass $m_{\D^{+(0)}\Lz}$ and three angular variables \mbox{$\Vec{\Omega}\equiv (\cos{\theta_{\Xires_c^{+(0)}}}, \phi_{\Lz}, \cos{\beta_{\Lz}})$}. 
The variable $\theta_{\Xires_c^{+(0)}}$ is the angle between the $\Lz$ momentum and the momentum of the pion from the $\Xires_b^{0(-)}$ decay, in the rest frame of the $\D^{+(0)}\Lz$ system (denoted as $\Xires_c^{+(0)}$), and is referred to as the $\Xires_c^{+(0)}$ helicity angle. 
Similarly, the $\Lz$ helicity angle $\beta_{\Lz}$ is defined by the momentum of the proton and that of the $\D^{+(0)}$ meson in the $\Lz$ rest frame.
The variable $\phi_{\Lz}$ is the angle between the ${\Xires_c^{+(0)}\to\D^{+(0)}\Lz}$ and ${\Lz\to\proton\pim}$ decay planes. 
These angles are illustrated in Fig.~\ref{fig:DecayAngles} in Appendix~\ref{sec:Appendix_for_formula}.
These variables are calculated by constraining, with a kinematic fit, the decay to originate from the PV and the $\Xires_b^{0(-)}$ mass to its known value. Their distributions are shown in Figs.~\ref{fig:ampfit_Xibz} and~\ref{fig:app_ampfit_Xibm} for the $\Xibz$ and $\Xibm$ channels, respectively, where the background is subtracted using the \sPlot weights.
The $\D^{+(0)}\Lz$ invariant-mass spectra clearly exhibit the
$\Xic(3055)^{+(0)}$ resonances, and also hint at the presence of the $\Xic(3080)^{+(0)}$ state as well as a nonresonant (NR) contribution.
This is the first observation of the $\Xic(3055)^{+(0)}$ baryons in $\Xires_b^{0(-)}$ decays.


Amplitude analyses for the $\Xibz$ and $\Xibm$ channels are carried out separately. 
In the following description, the notation applies to the $\Xibz$ channel, but is similar for the $\Xibm$ channel.
An unbinned maximum-likelihood fit is performed to the four-dimensional ${\Xibz\to\Dp\Lz\pim}$ distribution using an amplitude model of the $m_{\Dp\Lz}$ and $\Vec{\Omega}$ observables.
The fit model accounts for various $J^P$ hypotheses: \mbox{$J^P=1/2^{\pm}$, $3/2^{\pm}$, $5/2^{\pm}$} or $7/2^{\pm}$ for the $\Xic(3055)^+$ and $\Xic(3080)^+$ baryon, 
and $1/2^{\pm}$ for the nonresonant component. 
The combination of the $J^P$ hypotheses that gives the largest likelihood value is considered as the favored one.
The logarithm of the likelihood function ($\log\mathcal{L}$) is defined as
\begin{equation}
    \log{\mathcal{L}}(\Vec{\nu}) = 
        \frac{\sum_{i}  {w}_{i}}{\sum_{i}  {w}_{i}^2}\sum_{i} 
        {w}_{i}\times\log\left[ 
            \mathcal{P}(m^{i}_{\Dp\Lz},\Vec{\Omega}^i|\Vec{\nu})
        \right],\label{eq:likelihood}
\end{equation}
where $\mathcal{P}(m_{\Dp\Lz},\Vec{\Omega}|\Vec{\nu})$ is the signal probability density function (PDF), 
${w}_{i}$ is the signal \sPlot  weight~\cite{Xie:2009rka}, the index $i$ runs over the $\Xibz$ candidates in data,  and $\Vec{\nu}$ denotes the vector of free parameters.
The factor $\sum_{i} {w}_{i}/\sum_{i} {w}_{i}^2$ is applied for a correct determination of the fit parameter uncertainties~\cite{Langenbruch:2019nwe} in the presence of background.

The PDF is formed by the the squared amplitude summed over the helicities of the $\Xibz$ baryon, $\lambda_{\Xibz}$, and of the proton $\lambda_{\proton}$, as
\begin{align}
    \mathcal{P}(m_{\Dp\Lz},\Vec{\Omega}|\Vec{\nu}) = &
        \nonumber\frac{1}{I(\Vec{\nu})} 
        \sum_{\lambda_{\Xibz}, \lambda_{\proton}} 
        \left |
            \mathcal{M}_{\lambda_{\Xibz}, \lambda_{\proton}}(m_{\Dp\Lz},\Vec{\Omega}|\Vec{\nu})
        \right |^2
        \\&
        \times \Phi(m_{\Dp\Lz},\Vec{\Omega}) \epsilon(m_{\Dp\Lz},\Vec{\Omega}), 
\end{align}
where $\Phi(m_{\Dp\Lz},\Vec{\Omega})$ is the phase-space density function that depends on the final state kinematics, and $\epsilon(m_{\Dp\Lz},\Vec{\Omega})$ is the experimental efficiency that is evaluated with simulation.
The normalization is given by
\begin{align}
    I(\Vec{\nu}) \equiv &
        \nonumber \int 
        \sum_{\lambda_{\Xibz}, \lambda_{\proton}} 
        \left |
            \mathcal{M}_{\lambda_{\Xibz}, \lambda_{\proton}}(m_{\Dp\Lz},\Vec{\Omega}|\Vec{\nu})
        \right |^2\\
        &\times \Phi(m_{\Dp\Lz},\Vec{\Omega}) \epsilon(m_{\Dp\Lz},\Vec{\Omega}) \deriv m_{\Dp\Lz}\deriv\Vec{\Omega}, 
\label{eq:integral}
\end{align}
and is calculated numerically with a Monte Carlo integration method~\cite{1998AcNum...7....1C} utilizing simulated ${\Xires_b^{0}\to\Dp\Lz\pim}$ decays. 
The signal decay amplitude $\mathcal{M}_{\lambda_{\Xibz}, \lambda_{\proton}}(m_{\Dp\Lz},\Vec{\Omega}|\Vec{\nu})$ is constructed based on the helicity formalism~\cite{Jacob:1959at}, for which the full formula is described in detail in Appendix~\ref{sec:Appendix_for_formula}.
In the amplitude model, the invariant-mass distributions of the $\Xic(3055)^+$ and $\Xic(3080)^+$ resonances are each described by a relativistic Breit--Wigner function~\cite{PDG2024},
and that of the nonresonant component is described empirically by an exponential function. 
The mass and width of the $\Xic(3055)^{+}$ baryon are free parameters in the fit, while those of the $\Xic(3080)^+$ baryon are fixed to the known values~\cite{PDG2024}. The helicity couplings for each ${\decay{\Xibz}{\Xires_c^{+}\pim}}$ decay, $H_{\lambda_{\Xibz}}$, with $\lambda_{\Xibz}=\pm 1/2$ are free to vary. They are used to define the up-down asymmetry of the decay as
\begin{equation}
\alpha\equiv\frac{|H_{\lambda_{\Xibz}=+1/2}|^2-|H_{\lambda_{\Xibz}=-1/2}|^2}{|H_{\lambda_{\Xibz}=+1/2}|^2+|H_{\lambda_{\Xibz}=-1/2}|^2},
\end{equation}
for which a nonzero value indicates parity symmetry violation~\cite{Lee:1956qn}.
The helicity couplings for the ${\Lz\to\proton\pim}$ decay are fixed to the precise measurement obtained by \besiii~\cite{BESIII:2022qax}.

Among all the considered spin-parity assignments, the combination of \mbox{$J^P_{\Xic(3055)^+}=3/2^{+}$}, \mbox{$J^P_{\Xic(3080)^+}=5/2^{+}$}, and \mbox{$J^P_{\text{NR}}=1/2^-$} gives the largest maximum likelihood. 
The projections of the corresponding amplitude model are overlaid with data distributions in Fig.~\ref{fig:ampfit_Xibz}.
In this scenario, the $\Xic(3055)^+$ and $\Xic(3080)^+$ baryons are consistent with the two $D$-wave excitations of the $\Xicp$ flavor triplet, 
where the charm quark spin is antiparallel or parallel to the orbital angular momentum, respectively. 
The nonresonant component is consistent with an $S$-wave decay to the $\Dp\!\Lz$ final state.
The significance of the \mbox{$\Xires_b^{0}\to\Xic(3080)^{+}(\to\D^{+}\Lz)\pim$} signal is determined to be $4.4\sigma$, with a likelihood-ratio test considering amplitude models with or without the $\Xic(3080)^{+}$ contribution.
The branching fraction for the $\Xic(3080)^{+}$ baryon relative to that for the $\Xic(3055)^{+}$ baryon in the $\Xibz\to\Xires_c^{**+}(\to \Dp\Lz)\pim$ decay, denoted as $R_\BF$, is measured by comparing the integral of the PDF for the $\Xic(3080)^{+}$ component to that of the $\Xic(3055)^{+}$ baryon. 
The mass and width of the $\Xic(3055)^+$ baryon, and the up-down asymmetry of the $\decay{\Xibz}{\Xic(3055)^+\pim}$ decay are also measured, as summarized in Table~\ref{tab:result}.

According to theoretical calculations for a $\bar{3}_F$ beauty baryon decaying to a $\bar{3}_F$ charm baryon and a pseudoscalar via a color-allowed process, where factorization is expected to hold, the up-down asymmetry is close to $-1$~\cite{Cheng:1996cs,Chua:2018lfa,Li:2022hcn}.
The ${\Xibz\to\Xic(3055)^+\pim}$ decay is such a process in the case that $\Xic(3055)^+$ baryon is the $D$-wave $\lambda$-excitation.
Otherwise, if the $\Xic(3055)^+$ baryon is a $6_F$ state,  the up-down asymmetry could depart strongly from $-1$~\cite{Ke:2024aux}.
The up-down asymmetries measured under other $J^P_{\Xic(3055)^+}$ hypotheses are listed in Table~\ref{tab:NLL}. 
Under the favored hypothesis $J^P_{\Xic(3055)^+}=3/2^{+}$, $\alpha=-0.92\pm0.10\pm0.05$ is consistent with maximal parity violation, which is not the case for other $J^P$ assignments.

The same analysis is carried out for the ${\decay{\Xibm}{\Xic(3055)^0\pim}}$ channel. While the \mbox{$J^P_{\Xic(3080)^{0}}=5/2^{+}$} and $7/2^{+}$ hypotheses provide similar likelihood values, the result of \mbox{$J^P_{\Xic(3055)^{0}}=3/2^{+}$} is robust, despite the spin-parity of $\Xic(3080)^0$, and yields $\alpha=-0.92\pm0.16\pm0.22$ under \mbox{$J^P_{\Xic(3080)^{0}}=5/2^{+}$}, \mbox{$J^P_{\Xic(3055)^{0}}=3/2^{+}$}, \mbox{$J^P_{\text{NR}}=1/2^{-}$}. 
The significance of the \mbox{$\Xires_b^{-}\to\Xic(3080)^{0}(\to\D^{0}\Lz)\pim$} signal is determined to be $3.6\sigma$.
The corresponding fit projections on data distributions are shown in Fig.~\ref{fig:app_ampfit_Xibm}. 
The measured properties are also summarized in Table~\ref{tab:result}.

\begin{table}[!t]
    \renewcommand{\arraystretch}{1.2}
    \centering
    \caption{
        Measurement of the masses ($m$) and widths ($\Gamma$) for the $\Xic(3055)^{+(0)}$ baryons, the up-down asymmetries ($\alpha$) of the ${\Xires_b^{0(-)}\to\Xic(3055)^{+(0)}\pim}$ decays, and the relative branching fractions for $\Xic(3080)^{+(0)}$ and $\Xic(3055)^{+(0)}$ baryons ($R_\BF$). All results are obtained under the favored hypothesis $J^P_{\Xic(3055)^{+(0)}}=3/2^{+}$.
        The first uncertainties are statistical and the second are systematic.
        }
    \label{tab:result}
    \begin{tabular}{lr@{\:$\pm$\:}c@{\:$\pm$\:}lr@{\:$\pm$\:}c@{\:$\pm$\:}l}
    \hline
    Quantity            & \multicolumn{3}{c}{$\Xic(3055)^+$}    & \multicolumn{3}{c}{$\Xic(3055)^0$}\\\hline
    $m\,[\!\mevcc]$       & $3054.52$&$0.36$&$0.17$               & $3061.00$&$0.80$&$0.23$\\
    $\Gamma\,[\!\mevcc]$    & $8.01$&$0.76$&$0.34$                  & $12.4$&$2.0$&$1.1$ \\
    $\alpha$            & $-0.92$&$0.10$&$0.05$                 & $-0.92$&$0.16$&$0.22$\\
    $R_\BF$             & $0.045$&$0.023$&$0.006$               & $0.14$&$0.06$&$0.04$\\
    \hline
    \end{tabular}
\end{table}

\begin{figure}[!tb]
    \centering
    \includegraphics[width=0.48\columnwidth]{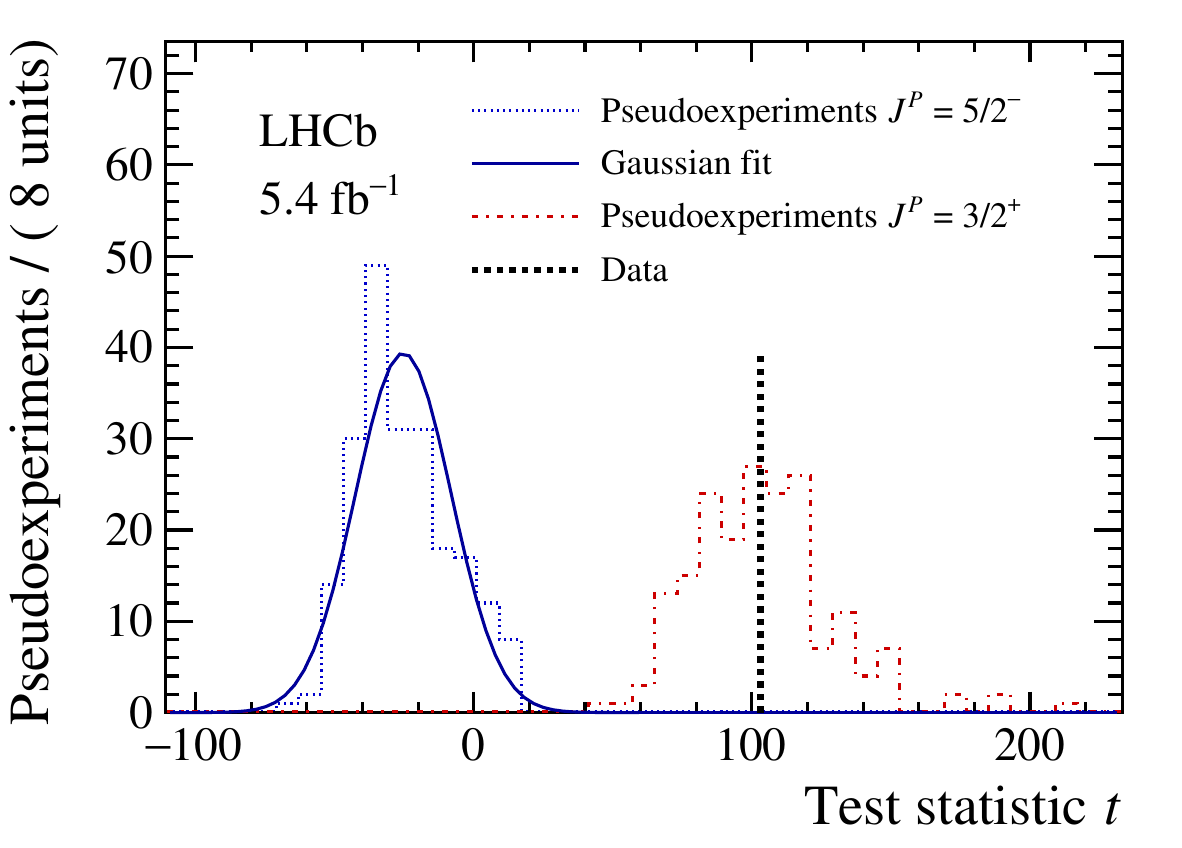}
    \includegraphics[width=0.48\columnwidth]{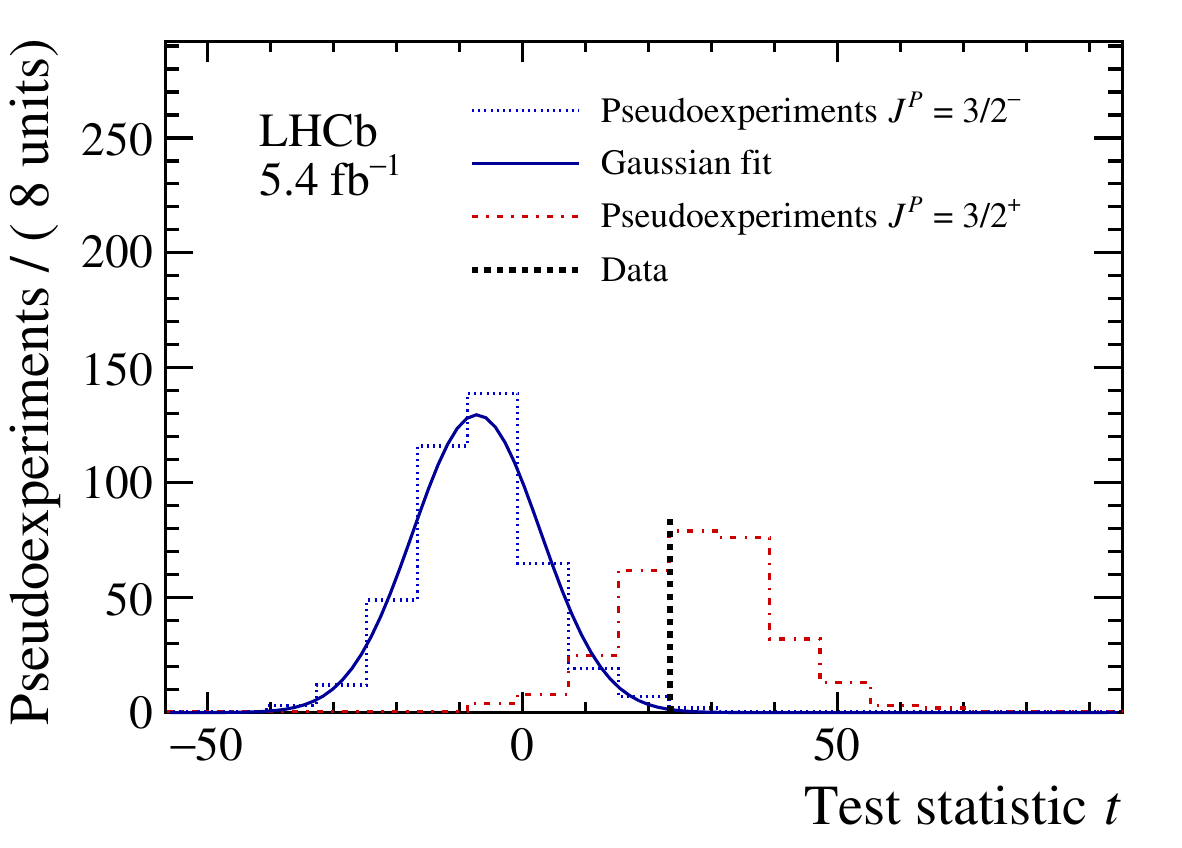}
    \caption{
    Distributions of the test statistic for (blue) pseudoexperiments generated with the alternative hypotheses of (left) ${J_{\text{alt}}^{P}=5/2^-}$ for the $\Xic(3055)^+$ baryon and (right)  ${J_{\text{alt}}^{P}=3/2^-}$ for the $\Xic(3055)^0$ baryon. Distributions for (red) pseudoexperiments generated with the $J^P=3/2^+$ hypothesis are also plotted as comparison. 
    The $J^P_{\Xic(3080)^{+(0)}}=5/2^{+}$ and $J^P_{\text{NR}}=1/2^{-}$ are fixed.
    Values of the test statistic in data are also given.
    }
    \label{fig:Significances}
\end{figure}

The significances of favoring $J^P_{\Xic(3055)^{+(0)}}=3/2^+$ over another hypothesis are evaluated with a likelihood-ratio test on pseudoexperiments.
Signal samples are generated with an alternative $J_{\text{alt}}^{P}$ for the $\Xic(3055)^{+(0)}$ baryons, along with background components.
The combined samples are then subjected to the same analysis procedure and fitted with $J^P=3/2^+$ and $J_{\text{alt}}^{P}$ hypotheses.
The difference between twice the $\log\mathcal{L}$ values of the two fits is taken as the test statistic $t$.
The $t$ distribution of pseudoexperiments is approximated by
a Gaussian function with $\mu( t _{J_{\text{alt}}^{P}})$ and $\sigma(t_{J_{\text{alt}}^{P}})$ as the mean and standard deviation, respectively.
The $J^P=3/2^+$ hypothesis is favored over the $J_\text{alt}^{P}$ hypothesis by a significance calculated as
\begin{equation}
    n_{\sigma}(J_{\text{alt}}^{P}) = \frac{t_{\text{data}} - \mu( t _{J_{\text{alt}}^{P}})}{\sigma(t_{J_{\text{alt}}^{P}})},
\end{equation}
where $t_{\text{data}}$ is the test statistic for data, calculated as twice of the likelihood difference between $J^P=3/2^+$ and $J_{\text{alt}}^{P}$ hypotheses in the data fitting.
Among all tested hypotheses, the minimum significance of the $J^P_{\Xic(3055)^{+(0)}}=3/2^+$ hypothesis is $6.5\sigma$ ($3.5\sigma$) against the \mbox{$J_{\text{alt}}^P = 5/2^-$} ($3/2^-$) hypothesis.
The $t$ distributions of pseudoexperiments generated with \mbox{$J^P_{\Xic(3055)^{+(0)}}=3/2^+$} and $5/2^-$ ($3/2^-$) are shown in Fig.~\ref{fig:Significances}, compared with the test statistic in data.
The significances of the $J^P_{\Xic(3055)^{+(0)}}=3/2^+$ hypothesis over all the tested alternative hypotheses are listed in Table~\ref{tab:NLL}. 
Given the $\Xic(3055)^+$ and $\Xic(3055)^0$ baryons are assumed to be isospin partners, the $J^P_{\Xic(3055)^{+(0)}}=3/2^+$ hypothesis is well established by these measurements.

\begin{table}[!tb]
\renewcommand{\arraystretch}{1.2}
\centering
\caption{
    Tested spin-parity hypotheses and the significance of favoring $J^P_{\Xic(3055)^{+(0)}}=3/2^+$ over each hypothesis, $n_{\sigma}$, where $J^P_{\Xic(3080)^{+(0)}}=5/2^{+}$ and $J^P_{\text{NR}}=1/2^{-}$ are fixed.
    Measured up-down asymmetries $\alpha$ in the ${\decay{\Xires_b^{0(-)}}{\Xic(3055)^{+(0)}\pim}}$ decays are also given, with statistical uncertainties. 
    }
\label{tab:NLL}
    \begin{tabular}{ccc}
    \hline
    $J^P_{\Xic(3055)^{+(0)}}$ & $n_{\sigma}$ & $\alpha$ \\\hline
    $1/2^-$         & $12.9\enspace(6.5)$                 & $ -0.10\pm0.17\enspace(-0.63\pm0.28)$ \\
    $1/2^+$         & $11.0\enspace(5.5)$                 & $ +0.31\pm0.13\enspace(+0.32\pm0.20)$ \\
    $3/2^-$         & $\phantom{0}7.3\enspace(3.5)$       & $ +0.18\pm0.14\enspace(+0.20\pm0.23)$ \\
    $5/2^-$         & $\phantom{0}6.5\enspace(4.8)$       & $ -0.12\pm0.14\enspace(-0.21\pm0.23)$ \\
    $5/2^+$         & $\phantom{0}9.8\enspace(4.8)$       & $ +0.52\pm0.14\enspace(+0.30\pm0.27)$ \\
    $7/2^-$         & $10.7\enspace(6.2)$                 & $ +0.41\pm0.16\enspace(+0.19\pm0.22)$ \\
    $7/2^+$         & $10.9\enspace(6.0)$                 & $ +0.12\pm0.14\enspace(-0.30\pm0.25)$ \\
    \hline
    \end{tabular}
\end{table}

Systematic uncertainties on the masses and widths of the $\Xic(3055)^{+(0)}$ baryons, the up-down asymmetries of the ${\decay{\Xires_b^{0(-)}}{\Xic(3055)^{+(0)}\pim}}$ transitions as well as the relative branching fractions $R_\BF$ are listed in Tables~\ref{tab:systematics} and~\ref{tab:systematics_Xic0} in Appendix~\ref{sec:Appendix_for_systematics}.
Possible biases introduced by the amplitude model are evaluated using pseudoexperiments, and are corrected for in the measurements. 
The uncertainties on the known masses of the $\Lz$, $\D^{+(0)}$ and $\pim$ hadrons~\cite{PDG2024} are propagated to the $\Xic(3055)^{+(0)}$ mass measurement.
Momentum-scale calibration for charged particles yields an uncertainty on the $\Xic(3055)^{+(0)}$ mass measurements~\cite{LHCb-DP-2023-003}. 
The experimental resolution smears the $m_{\D^{+(0)}\Lz}$ invariant-mass distribution, and introduces an uncertainty. 
The limited size of the simulation sample used for the PDF normalization results in an uncertainty, which is evaluated using the bootstrap method~\cite{efron:1979}.
Corrections are applied to simulation to match the trigger performances in data~\cite{LHCb-PAPER-2022-038}, and the uncertainty on this correction leads to a systematic uncertainty. 
The $\Xires_b^{0(-)}$ candidates are reconstructed in two categories, depending on whether the $\Lz$ baryon decays within or downstream of the \lhcb vertex detector~\cite{LHCb-DP-2014-001}. The possible efficiency difference between them is studied as a source of systematic uncertainty.
In the fit to the invariant-mass distribution of the $\Xires_b^{0(-)}$ baryon, the models for the signal, the combinatorial and the partially reconstructed background are varied to evaluate the corresponding systematic uncertainty.
The orbital angular momentum between the $\Xires_c^{+(0)}$ baryon and the $\pim$ meson in the ${\decay{\Xires_b^{0(-)}}{\Xires_c^{+(0)}\pim}}$ decay is fixed to the lowest value in the baseline amplitude fit, and a systematic uncertainty is evaluated by considering all possible values.
An exponential function is used to describe the nonresonant invariant-mass distribution. An alternative linear function is tested, and the difference from the baseline result is taken as an uncertainty.
The fixed $\Xic(3080)^{+(0)}$ masses and widths are varied within their uncertainties, resulting in two sources of systematic uncertainties.
It is possible that the same track segment is shared by more than one track in the $\Xires_b^{0(-)}$ final state, resulting in cloned tracks. The results obtained by removing such candidates are compared with the baseline results and the differences are quoted as systematic uncertainties. 
The overall systematic uncertainties are obtained with quadratic sum of the contributions, and are comparable to the statistical uncertainties.

In conclusion, the ${\Xires_b^{0(-)}\to\Xic(3055)^{+(0)}\pim}$ decays, with ${\Xic(3055)^{+(0)}\to D^{+(0)}\Lz}$ are observed for the first time in $pp$ collisions using data recorded at $\sqrt{s}=13\tev$, corresponding to an integrated luminosity of $5.4\invfb$.
An amplitude analysis is performed on each channel independently, determining for the first time the spin-parity of the $\Xic(3055)^{+(0)}$ baryons to be $3/2^+$, with significances of more than $6.5\sigma$ ($3.5\sigma$) against other hypotheses. Different sources of systematic uncertainties have been taken into account, confirming the conclusions of the spin-parity assignment.
With the spin-parity assignment of $J^P=3/2^+$, the up-down asymmetries of the ${\Xires_b^{0(-)}\to\Xic(3055)^{+(0)}\pim}$ decays are measured to be $-0.92\pm0.10\stat\pm0.05\syst$ ($-0.92\pm0.16\stat\pm0.22\syst$), consistent with maximal parity violation.
This is the first measurement of the parity-violating parameter for the transition of the $\Xires_b^{0(-)}$ baryon to a $\Xires_c^{+(0)}$ baryon and a pseudoscalar meson.
The result supports the factorization approximation in color-allowed beauty-to-charm baryon decays, which indicates the structure of the $\Xic(3055)^{+(0)}$ state. 
The masses and widths of the $\Xic(3055)^{+(0)}$ baryons are also measured, with a precision comparable to known results~\cite{PDG2024}.
All the obtained results for the $\Xic(3055)^{+(0)}$ state support its interpretation as the first $D$-wave excitation of the flavor antisymmetric $\bar{3}_F$ $\Xic$ state. 
The significances of the \mbox{$\Xires_b^{0(-)}\to\Xic(3080)^{+(0)}(\to\D^{+(0)}\Lz)\pim$} signal are determined to be $4.4\sigma$ ($3.6\sigma$), and their branching fractions relative to those of the \mbox{${\Xires_b^{0(-)}\to\Xic(3055)^{+(0)}(\to\D^{+(0)}\Lz)\pim}$} decays are measured for the first time.

%% file: acknowledgements.tex
\section*{Acknowledgements}
%
% These Acknowledgements valid from 3-May-2019
%
\noindent We express our gratitude to our colleagues in the CERN
accelerator departments for the excellent performance of the LHC. We
thank the technical and administrative staff at the LHCb
institutes.
We acknowledge support from CERN and from the national agencies:
CAPES, CNPq, FAPERJ and FINEP (Brazil); 
MOST and NSFC (China); 
CNRS/IN2P3 (France); 
BMBF, DFG and MPG (Germany); 
INFN (Italy); 
NWO (Netherlands); 
MNiSW and NCN (Poland); 
MCID/IFA (Romania); 
%MSHE (Russia); 
MICIU and AEI (Spain);
SNSF and SER (Switzerland); 
NASU (Ukraine); 
STFC (United Kingdom); 
DOE NP and NSF (USA).
We acknowledge the computing resources that are provided by CERN, IN2P3
(France), KIT and DESY (Germany), INFN (Italy), SURF (Netherlands),
PIC (Spain), GridPP (United Kingdom), 
%RRCKI and Yandex LLC (Russia), 
CSCS (Switzerland), IFIN-HH (Romania), CBPF (Brazil),
and Polish WLCG (Poland).
We are indebted to the communities behind the multiple open-source
software packages on which we depend.
Individual groups or members have received support from
ARC and ARDC (Australia);
Key Research Program of Frontier Sciences of CAS, CAS PIFI, CAS CCEPP, 
Fundamental Research Funds for the Central Universities, 
and Sci. \& Tech. Program of Guangzhou (China);
Minciencias (Colombia);
EPLANET, Marie Sk\l{}odowska-Curie Actions, ERC and NextGenerationEU (European Union);
A*MIDEX, ANR, IPhU and Labex P2IO, and R\'{e}gion Auvergne-Rh\^{o}ne-Alpes (France);
%RFBR, RSF and Yandex LLC (Russia);
AvH Foundation (Germany);
ICSC (Italy); 
%GVA, XuntaGal, GENCAT, Inditex, InTalent and Prog.~Atracci\'on Talento, CM (Spain);
Severo Ochoa and Mar\'ia de Maeztu Units of Excellence, GVA, XuntaGal, GENCAT, InTalent-Inditex and Prog. ~Atracci\'on Talento CM (Spain);
SRC (Sweden);
the Leverhulme Trust, the Royal Society
 and UKRI (United Kingdom).

%% file: supplemental-journal.tex
\clearpage

\section*{End matter}

\section{Formula of the amplitude model}
\label{sec:Appendix_for_formula}

\def\Bdecay{\Xib}
\def\Cdecay{\Xic\rightarrow \D\Lz}
\def\Ldecay{\Lz}
\def\BdecayNR{{\Xib\rightarrow \text{NR} {\pi}}}
\def\NRdecay{\text{NR}\rightarrow \D\Lz}

\def\lb{\lambda_{\Xib}}
\def\lc{\lambda_{\Xic}}
\def\lpip{\lambda_{\pim}} %from Xib
\def\ll{\lambda_{\Lz}}
\def\ld{\lambda_\D}
\def\lp{\lambda_\proton}
\def\lpim{\lambda_{\pim}} %from Lambda

\def\myJ{{J}}
\def\myP{{P}}
\def\mytheta{   {\theta_{\Xic}} }
\def\myphi{     {\phi_{\Lz}  } }
\def\mybeta{    {\beta_{\Lz}    } }
\def\mygamma{   {\gamma   } }

\newcommand{\dfunc}[4]
{ d^{#1}_{{#2},{#3}}({#4}) }

The full formula of the amplitude model used to describe $\decay{\Xib}{\Xires_c\pim}$ decays is 
%\begin{widetext}
\begin{align}
    \mathcal{M}_{\lb,\lp}(m_{\D\Lz},\Vec{\Omega}|\Vec{\nu}) = 
    &\nonumber
        \sum_{\Xires_c} 
        \sum_{\ll=\pm\frac{1}{2}} 
        (-1)^{J+1/2} 
        \times P 
        \times H^{\Bdecay}_{\lb} 
        \times H^{\Ldecay}_{\lp}
        \times
        \\\nonumber & 
        \dfunc{\myJ}{\lb}{\ll}{\mytheta} \dfunc{1/2}{\ll}{\lp}{\mybeta} e^{i\myphi\ll} 
        \ R(m_{\D\Lz})
    ,
\label{equ:total}
\end{align}
%\end{widetext}
where the $\Xic$ refers to the $\Xic(3055)$ and $\Xic(3080)$ resonances, as well as the nonresonant component, 
$\lambda$ is the helicity of a given particle, 
and $H$ is the helicity coupling of the corresponding decay. 
The $\decay{\Xires_c}{\D\Lz}$ strong decay contributes to a $(-1)^{J+1/2}$ term, assuming parity is conserved in such decays.
Finally, $J$ and $P$ are the spin and parity of the $\Xic$ baryon, 
$d$ is the Wigner small-$d$ function, and the $R(m_{\D\Lz})$ terms are Breit--Wigner functions convolved with a Gaussian function for the $\Xic$ signal, and exponential functions for nonresonant background.

\begin{figure}[!hb]
\centering
\includegraphics[width=0.96\columnwidth]{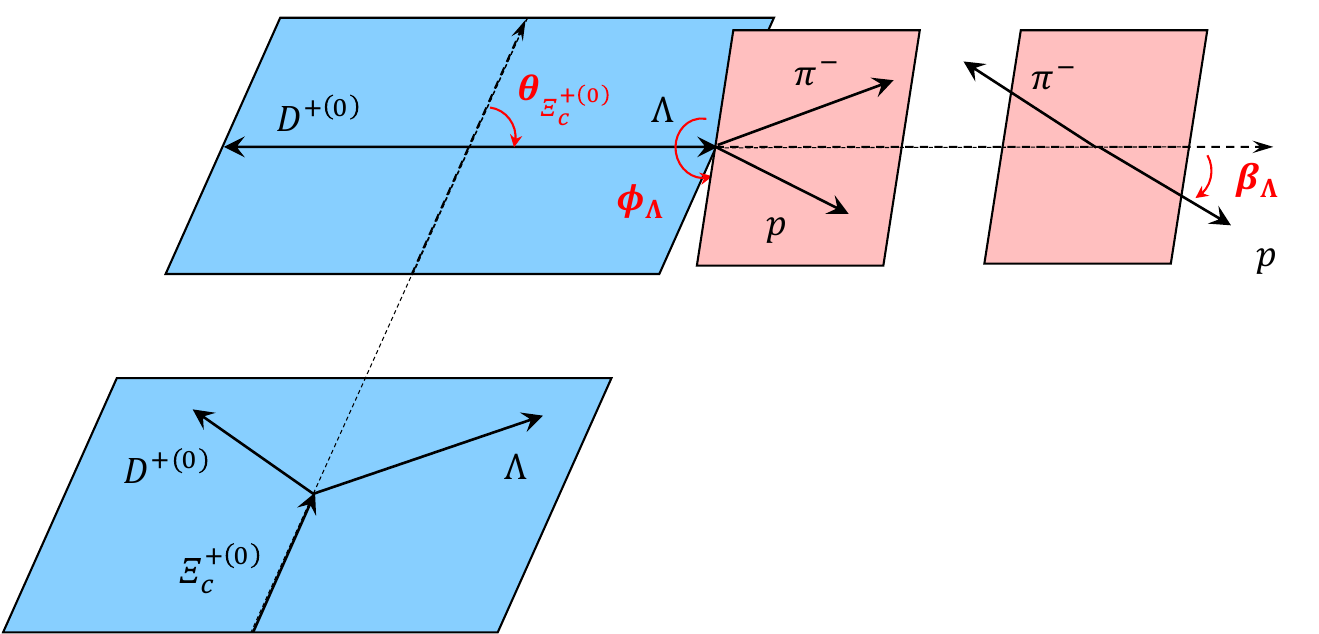}
\caption{Definition of the three angular variables \mbox{$\cos{\theta_{\Xires_c}}, \phi_{\Lz}$ and $ \cos{\beta_{\Lz}}$}. The variable $\theta_{\Xires_c^{+(0)}}$ is the angle between the $\Lz$ momentum and the momentum of the pion from the $\Xires_b^{0(-)}$ decay, in the rest frame of the $\D^{+(0)}\Lz$ system (denoted as $\Xires_c^{+(0)}$), and is referred to as the $\Xires_c^{+(0)}$ helicity angle. The $\Lz$ helicity angle $\beta_{\Lz}$ is defined by the momentum of the proton and that of the $\D^{+(0)}$ meson in the $\Lz$ rest frame. The variable $\phi_{\Lz}$ is the angle between the ${\Xires_c^{+(0)}\to\D^{+(0)}\Lz}$ and ${\Lz\to\proton\pim}$ decay planes. }
\label{fig:DecayAngles}
\end{figure}

The definition of the decay angles is illustrated in Fig.~\ref{fig:DecayAngles}.

\clearpage
\section{Summary table of systematic uncertainties}
\label{sec:Appendix_for_systematics}

The biases and systematic uncertainties on the masses ($m$) and widths ($\Gamma$) for the $\Xic(3055)^{+(0)}$ baryons, the up-down asymmetries ($\alpha$) of the ${\Xires_b^{0(-)}\to\Xic(3055)^{+(0)}\pim}$ decays, and the relative branching fractions for $\Xic(3080)^{+(0)}$ and $\Xic(3055)^{+(0)}$ baryons ($R_\BF$) are summarized in Tables~\ref{tab:systematics} and~\ref{tab:systematics_Xic0}.

\vfill
\begin{table}[!hb]
    \centering
    \caption{
    Biases and systematic uncertainties for the ${\Xires_b^{0}\to\Xic(3055)^{+}\pim}$ channel.
    }
    \begin{tabular}{lcccc}
        %%%%%%%%%%%%%%%%%%%%%%%%%%%%%%%%%%%%%%%%%%%%%%%%
        \hline \textbf{Source}
                                &$\sigma_{m}\,[\!\mevcc]$   &$\sigma_{\Gamma}\,[\!\mevcc]$  &$\sigma_{\alpha}$   &$\sigma_{R_\BF}$\\\hline
        %%%%%%%%%%%%%%%%%%%%%%%%%%%%%%%%%%%%%%%%%%%%%%%%
        Amplitude fit bias      &$-$              &$-$                 &$-$                        &$-$\\
        Hadron masses           &$\pm0.05$        &$-$                 &$-$                        &$-$\\
        Momentum scale          &$\pm0.01$        &$-$                 &$-$                        &$-$\\
        Resolution              &$\pm0.00$        &$\pm0.07$            &$\pm0.00$            &$\pm0.000$\\
        Simulation sample       &$\pm0.15$        &$\pm0.30$            &$\pm0.02$            &$\pm0.002$\\
        Trigger correction      &$\pm0.01$        &$\pm0.03$            &$\pm0.02$            &$\pm0.000$\\
        $\Lz$ categories        &$\pm0.03$        &$\pm0.04$            &$\pm0.01$            &$\pm0.002$\\
        $\Xibz$ mass fit model  &$\pm0.03$        &$\pm0.13$            &$\pm0.01$            &$\pm0.001$\\
        Angular momentum        &$\pm0.00$        &$\pm0.00$            &$\pm0.04$            &$\pm0.002$\\
        Nonresonant model       &$\pm0.00$  &$\pm0.00$            &$\pm0.00$            &$\pm0.000$\\
        $\Xic(3080)^{+}$ width  &$\pm0.01$        &$\pm0.01$            &$\pm0.00$            &$\pm0.003$\\
        $\Xic(3080)^{+}$ mass   &$\pm0.00$        &$\pm0.02$            &$\pm0.00$            &$\pm0.000$\\
        Clone tracks            &$\pm0.02$        &$\pm0.03$            &$\pm0.01$            &$\pm0.003$\\

        %%%%%%%%%%%%%%%%%%%%%%%%%%%%%%%%%%%%%%%%%%%%%%%%
        \hline\textbf{Total}
                                &$\pm0.17$        &$\pm0.34$            &$\pm0.05$            &$\pm0.006$\\\hline
        %%%%%%%%%%%%%%%%%%%%%%%%%%%%%%%%%%%%%%%%%%%%%%%%
    \end{tabular}
    \label{tab:systematics}
\end{table}

\begin{table}[!hb]
    \centering
    \caption{
    Biases and systematic uncertainties for the ${\Xires_b^{-}\to\Xic(3055)^{0}\pim}$ channel.
    }
    \begin{tabular}{lcccc}
        %%%%%%%%%%%%%%%%%%%%%%%%%%%%%%%%%%%%%%%%%%%%%%%%
        \hline \textbf{Source}
                                &$\sigma_{m}[\!\mevcc]$   &$\sigma_{\Gamma}[\!\mevcc]$  &$\sigma_{\alpha}$   &$\sigma_{R_\BF}$\\\hline
        %%%%%%%%%%%%%%%%%%%%%%%%%%%%%%%%%%%%%%%%%%%%%%%%
        Amplitude fit bias      &$-$              &$-0.46$             &$-$                        &$-$\\
        Hadron masses           &$\pm0.05$        &$-$                 &$-$                        &$-$\\
        Momentum scale          &$\pm0.03$        &$-$                 &$-$                        &$-$\\
        Resolution              &$\pm0.00$        &$\pm0.10$            &$\pm0.00$            &$\pm0.001$\\
        Simulation sample       &$\pm0.13$        &$\pm0.38$            &$\pm0.02$            &$\pm0.006$\\
        Trigger correction      &$\pm0.01$        &$\pm0.03$            &$\pm0.00$            &$\pm0.001$\\
        $\Lz$ categories        &$\pm0.04$        &$\pm0.12$            &$\pm0.05$            &$\pm0.004$\\
        $\Xibm$ mass fit model  &$\pm0.00$        &$\pm0.19$            &$\pm0.02$            &$\pm0.003$\\
        Angular momentum        &$\pm0.01$        &$\pm0.15$            &$\pm0.21$            &$\pm0.014$\\
        Nonresonant model       &$\pm0.00$        &$\pm0.03$            &$\pm0.00$            &$\pm0.001$\\
        $\Xic(3080)^{0}$ width  &$\pm0.08$        &$\pm0.69$            &$\pm0.01$            &$\pm0.032$\\
        $\Xic(3080)^{0}$ mass   &$\pm0.03$        &$\pm0.20$            &$\pm0.01$            &$\pm0.006$\\
        Clone tracks            &$\pm0.13$        &$\pm0.04$            &$\pm0.04$            &$\pm0.008$\\
        %%%%%%%%%%%%%%%%%%%%%%%%%%%%%%%%%%%%%%%%%%%%%%%%
        \hline\textbf{Total}
                                &$\pm0.23$        &$\pm1.11$            &$\pm0.22$            &$\pm0.038$\\\hline
        %%%%%%%%%%%%%%%%%%%%%%%%%%%%%%%%%%%%%%%%%%%%%%%%
    \end{tabular}
    \label{tab:systematics_Xic0}
\end{table}

\clearpage

%% file: Authorship_LHCb-PAPER-2024-018.tex
% LHCb collaboration author list
% Data extracted on August 30th, 2024 at 11:35am for paper reference LHCb-PAPER-2024-018
\centerline
{\large\bf LHCb collaboration}
\begin
{flushleft}
\small
R.~Aaij$^{36}$\lhcborcid{0000-0003-0533-1952},
A.S.W.~Abdelmotteleb$^{55}$\lhcborcid{0000-0001-7905-0542},
C.~Abellan~Beteta$^{49}$,
F.~Abudin{\'e}n$^{55}$\lhcborcid{0000-0002-6737-3528},
T.~Ackernley$^{59}$\lhcborcid{0000-0002-5951-3498},
A. A. ~Adefisoye$^{67}$\lhcborcid{0000-0003-2448-1550},
B.~Adeva$^{45}$\lhcborcid{0000-0001-9756-3712},
M.~Adinolfi$^{53}$\lhcborcid{0000-0002-1326-1264},
P.~Adlarson$^{80}$\lhcborcid{0000-0001-6280-3851},
C.~Agapopoulou$^{13}$\lhcborcid{0000-0002-2368-0147}
% ...

{\footnotesize \it

% ...
$^{13}$Universit{\'e} Paris-Saclay, CNRS/IN2P3, IJCLab, Orsay, France\\
% ...
$^{36}$Nikhef National Institute for Subatomic Physics, Amsterdam, Netherlands\\
% ...
$^{45}$Instituto Galego de F{\'\i}sica de Altas Enerx{\'\i}as (IGFAE), Universidade de Santiago de Compostela, Santiago de Compostela, Spain\\
% ...
$^{49}$Physik-Institut, Universit{\"a}t Z{\"u}rich, Z{\"u}rich, Switzerland\\
% ...
$^{53}$H.H. Wills Physics Laboratory, University of Bristol, Bristol, United Kingdom\\
% ...
$^{55}$Department of Physics, University of Warwick, Coventry, United Kingdom\\
% ...
$^{58}$School of Physics and Astronomy, University of Glasgow, Glasgow, United Kingdom\\
$^{59}$Oliver Lodge Laboratory, University of Liverpool, Liverpool, United Kingdom\\
% ...
$^{67}$Syracuse University, Syracuse, NY, United States\\
% ...
$^{80}$Department of Physics and Astronomy, Uppsala University, Uppsala, Sweden, associated to $^{58}$\\
% ...
}
\end{flushleft}